\documentclass{aa}  

\usepackage{graphicx}
\usepackage{longtable}
\usepackage{tabularx, blindtext}
\usepackage{lscape}
\usepackage[varg]{txfonts}
\usepackage{lmodern}
\begin{document}

   \title{Sulfur abundances in the Galactic bulge and disk\thanks{This paper is based on data collected with the Very Large Telescope (VLT) at the European Southern Observatory (ESO) on Paranal, Chile (ESO Program ID 065.L-0507, 071.B-0529, 076.B-0055, 076.B-0133, 077.B-0507, 079.D-0567, 082.B-0453, 083.B-0265, 084.B-0837, 084.D-0965, 085.B-0399, 086.B-0757, 087.B-0600, 087.D-0724, 088.B-0349, 089.B-0047, 090.B-0204, 091.B-0289, 092.B-0626, 093.B-0700, 094.B-0282, 165.L-0263, 167.D-0173, 266.D-5655; and data from the UVES Paranal Observatory Project (ESO DDT Program ID 266.D-5655).}}

   \author{F. Lucertini\inst{1}, L. Monaco\inst{1},
   E. Caffau\inst{2}, P. Bonifacio\inst{2}, A. Mucciarelli\inst{3,4}}

   \institute{Departamento de Ciencias Fisicas, Universidad Andres Bello, Fernandez Concha 700, Las Condes, Santiago, Chile
   \and
   GEPI, Observatoire de Paris, Universit\`e PSL, CNRS,
Place Jules Janssen, 92195 Meudon, France
	\and 
	Dipartimento di Fisica e Astronomia, Università degli Studi di Bologna, Via Gobetti 93/2, I-40129 Bologna, Italy
	\and	
	INAF - Osservatorio di Astrofisica e Scienza dello Spazio di Bologna, Via Gobetti 93/3, I-40129 Bologna, Italy}

   \date{Received 2021 March 13; Accepted 2021 September 9}

 
  \abstract
   {The measurement of $\alpha$-elements abundances provides a powerful tool to put constraints on chemical evolution and star formation history of galaxies.   
    The majority of studies on the $\alpha$-element Sulfur (S) are focused on local stars, making S behavior in other 
environments an astronomical topic yet to be analyzed.}
   {The investigation of S in the Galactic bulge has only recently been considered for the first time. This work aims to improve our knowledge on S behavior in this component of the Milky Way.}
   {We present S abundances of 74 dwarf and sub-giant stars in the Galactic bulge, 21 and 30 F and G thick and thin disk stars. We performed local thermodynamic equilibrium analysis and applied corrections for non-LTE on high resolution and high signal-to-noise UVES spectra.
   S abundances were derived from multiplets 1, 6 and 8 in the metallicity range $-2<$[Fe/H]$<$0.6, by spectrosynthesis or line equivalent widths.}
   {We confirm that S behaves like an $\alpha$-element within the Galactic bulge. In the [S/Fe] versus [Fe/H] diagram, S presents a plateau at low metallicity followed by a decreasing of [S/Fe] with the increasing of [Fe/H], until reaching [S/Fe]$\sim0$ at super-solar metallicity. We found that the Galactic bulge is S-rich with respect to both the thick and thin disks at $-1<$[Fe/H]$<0.3$, supporting a more rapid formation and chemical evolution of the Galactic bulge than the disk.}
   {}

   \keywords{Stars: abundances -- Galaxy: bulge -- Galaxy: disk}
   \authorrunning{F. Lucertini et al.}
   \maketitle
   
%

\section{Introduction}
The Galactic bulge (bulge) is one of the main, more massive and older components of the Milky Way (MW). 
Despite the spectroscopic 
\citep{1994ApJS...91..749M,2007ApJ...658L..29R, 2012EPJWC..1906003N, 2014A&A...569A.103R, 2014AJ....148...67J, 2017A&A...599A..12Z} and photometric \citep{2008ApJ...684.1110C, 2015ASPC..491..182G, 2018ApJ...863...16R} studies done in the last thirty years, 
the formation and the evolution of bulge is still an  open issue. 
\cite{2004ARA&A..42..603K} classified local bulges into classical and pseudo-bulges.
From the morphological point of view, classical bulges are more spheroidal and massive with respect to disks. Their shape suggests that they were created from gravitational collapse of primordial gas (\citealt{1962ApJ...136..748E}, \citealt{1990ApJ...365..539M}), or by hierarchical merging of smaller structures (\citealt{1984Natur.311..517B}). According to this theory, classical bulges formed during the first phase of galaxy formation, resulting older than disks.
On the other hand, pseudo-bulges are associated to bars and smaller disk-like structures.
Their origin is due to the secular evolution of disks. In particular, the internal instabilities of the disk bring gas and stars into the central regions of the galaxy, building up the bulge (\citealt{2005MNRAS.358.1477A}, \citealt{2008ApJ...687...59G}). 
Nowadays, the idea that more than one scenario are involved into the formation of bulges is strongly supported \citep{2006ApJ...645..986R, 2009ApJ...707L...1B, 2010MNRAS.401.1826R}.

Since the Galactic bulge is the only one 
that can be resolved into individual stars, it represents a unique 
laboratory to understand the formation and evolution of bulges and their galaxies.
The analysis of bulge stars contributes to put constrains on the formation timescales and chemical enrichment of the earliest stellar populations in the galaxy.
In particular, the chemical evolution and the star formation history (SFH) of the bulge 
can be derived from  measured abundance ratios.

The $\alpha$-elements and iron-peak elements 
play a fundamental role in this context since their nucleosynthesis 
occurs on different timescales.
While $\alpha$-elements are produced by Type II Supernovae (SNe) on a timescale of 
$\sim$30 Myr, 
iron-peak elements are produced by
Type Ia SNe on longer timescales ($>$1-2 Gyr).
Thus, 
the investigation of the $\alpha$-elements behavior in the bulge 
is a crucial topic to understand its formation and evolution.

\cite{2005ApJ...634.1293R} measured $\alpha$-elements (O, Mg, Si, Ca, Ti) abundances in 14 M giant stars belonging to the bulge in the metallicity range $-0.35<$[Fe/H]$<0$ using infra-red spectra.
Overall, they found an $\alpha$-enhancement of $\sim +0.3$ dex in bulge stars with respect to local disk. 
Using high resolution UVES spectra, \cite{2008MmSAI..79..503Z} measured  O, Na, Mg and Al abundances for 50 K giants in the bulge. 
In the metallicity range $-0.8<$[Fe/H]$<0.4$, they obtained [O/Fe] and [Mg/Fe] bulge trends higher than those measured in both thin and thick disk stars.
Taking advantage of microlensing events, in a series of paper, Bensby and collaborators (\citeyear{2013A&A...549A.147B}, \citeyear{2017A&A...605A..89B}, \citeyear{2020A&A...634A.130B}) spectroscopically characterized the bulge studying dwarf and sub-giant stars.
Among the 13 species considered, they measured Mg, Si, Ca and Ti abundances and they found that the level of $\alpha$-enhancement is slightly higher in the bulge than in the thick disk. Moreover, the bulge presents a knee in the [$\alpha$/Fe] trend located at $\sim$0.1 dex higher [Fe/H] than the local thick disk.
It should be kept in mind that while all the stars studied by Bensby
and collaborators lie in the direction of the bulge, 
they are not necessarily well established members of the bulge.
Thus, all these works conclude that the bulge formed before and more rapidly than the disk. 

On the other hand, \cite{2008A&A...484L..21M} measured O abundances from high resolution NIR spectra of 19 bulge giant stars and found no chemical distinction between the bulge and the local thick disk at [Fe/H]$<-0.2$.
They analyzed the formation timescales, star formation rate and initial mass function of bulge and disk to argue the similar chemical evolution of these components of the MW.
Similar results were obtained for three K bulge giant stars by \cite{2009A&A...496..701R}. According to this work, the O abundances measured from high-resolution NIR spectra in bulge and thick disk stars are similar.
The similarity between the bulge and thick disk trends was also confirmed by \cite{2011A&A...530A..54G, 2015A&A...584A..46G}, who measured Mg, Ca, Ti and Si abundances from high resolution (R) and high signal-to-noise (S/N) FLAMES/GIRAFFE spectra of bulge giants.

Recently, \cite{2020arXiv200905063G} investigated the similarity or dissimilarity between the chemical trends of Apache Point Observatory Galactic Evolution Experiment (APOGEE, \citealt{2017AJ....154...94M}) bulge (R$_{GC}<$3 kpc) and thick disk (5 kpc$<R_{GC}<$11 kpc) stars.
Dividing bulge and thick disk stars in low-Ia (high-[Mg/Fe]) and high-Ia (low-[Mg/Fe]) populations, they measured chemical abundances of seventeen species to analyze the median trend in the [X/Mg] vs [Mg/H] plane.
They found nearly identical median trends for low-Ia thick disk and bulge stars for all elements, while the high-Ia trends are similar for most elements except Mn, Na and Co.
Obtaining typical differences $\lesssim 0.03$ dex between the abundance trends, \cite{2020arXiv200905063G} concluded that bulge and thick disk were enriched by similar nucleosynthetic processes.

Among the different $\alpha$-elements, Sulfur has not been thoroughly studied.
Sulfur (S) is produced in the final stage of the evolution of massive stars (M$_*>20$M$\odot$). The hydrostatic burning of neon (Ne), at the core temperature $logT_c=9.09 $ K, leads to the formation of an oxygen (O) convective core and the production of $\alpha$-elements up to $^{32}$S. Once $logT_c=9.5 $ K is reached, the oxygen burning phase starts producing $^{28}$Si, $^{32}$S and $^{34}$S. This amount of Sulfur will be almost completely destroyed during the Si burning phase at 2.3x10$^{9}$K, so it will not contribute to the gas forming the next generation of stars. However, the production of S continues in the O convective shell burning in upper layers and it is provided also by explosive O burning during type II SNe explosion \citep{2003ApJ...592..404L, 2020ApJ...900..179K}.
Unlike other $\alpha$-elements, S is moderately volatile. For this reason, its abundance measured in stars in the Local Group galaxies can be directly compared to those measured in extra-galactic HII regions or damped Ly-$\alpha$ systems (DLA). 
Hence, S abundances (A(S)) provide clues on the SFH and on properties of the interstellar medium (ISM), connecting the local and distant Universe.
In stellar spectra a few S multiplets (Mult.) are available.
In the optical range there are Mult. 1, 6 and 8 at 920 nm, 870 nm and 675 nm, respectively.
Then, the Mult. 3 at 1045 nm and a forbidden line [SI] (1082 nm) lies in the near-infrared (NIR) part of the spectra.
Mult. 6 and 8 features are useful for solar metallicity or slightly metal-poor stars.
However, these lines are weak so high resolution and high S/N spectra are required to use them.
The strongest features of S are those of Mult. 1 and 3, allowing to investigate A(S) in the metal-poor regime. The analysis of solar-metallicity dwarf stars and metal-poor giants can be carried out from the line [SI] at 1082 nm.
On the other hand, these lines are located in a spectral range strongly contaminated by telluric  absorptions.
For these reasons, the study of S is often omitted 
in favor of the easier analysis of other $\alpha$-elements.
As a result 
our knowledge about the behavior of S is still poor with respect to that of
other $\alpha$-elements.

This is particularly true when we consider the Galactic bulge. Indeed, A(S) have only recently been measured for the first time in this component of the MW  by \cite{2020arXiv200905063G} using the \ion{S}{i} lines at 15478.5 $\AA$ and 16576.6 $\AA$, measured on high resolution NIR APOGEE spectra of red giant branch (RGB) stars.
In this work, we present new S abundances in dwarf and sub-giant branch (SGB) stars belonging to the bulge and the disk based on high resolution, high S/N archival optical spectra, using the multiplets 1, 6 and 8 and we compare ours with literature results.

The paper is structured as follows: the used data sets are described in Section 2.
In Section 3 we explain the analysis and the method adopted to obtain the A(S).
We compare our results with those in literature in Section 4. Finally, conclusion follows in Section 5.

\section{Observational data} 
Considering the distance of the bulge and the high degree of interstellar extinction in the Galactic plane, giant stars are the best tracers for the bulge.
However, all the S lines we use are of high excitation so they become weak in cool stars.
For this reason, dwarf stars can be considered among the best tracers of Galactic chemical evolution of S.

The faintness of bulge dwarf stars is the main difficulty in their
observation. 
High-resolution spectra of these stars can be obtained when a gravitational micro-lensing event occurs, during which the star may brighten by factors of hundreds.
Several bulge dwarfs and sub-giant stars have been observed during microlensing events (\citealt{2009A&A...499..737B}, \citeyear{2010A&A...521L..57B}, \citeyear{2010A&A...512A..41B}, \citeyear{2011A&A...533A.134B}, \citeyear{2013A&A...549A.147B}, \citeyear{2017A&A...605A..89B}, \citeyear{2020A&A...634A.130B}).
On these occasions, from 2 to 8 $\sim$1800 s UVES (\citealt{2000SPIE.4008..534D}) exposures were recorded with a 0.7$''$ wide slit 
with wavelength coverage in the intervals 376-498 nm, 568-750nm and 766-946 nm.
During the same observational night, one rapidly rotating B star for each target was observed to evaluate the effect due to the Earth atmosphere.
We retrieved the UVES reduced (UVES pipeline version 5.10.13\footnote{https://www.eso.org/sci/software/pipelines/}) data from the ESO data archive\footnote{http://archive.eso.org/wdb/wdb/adp/phase3$\_$spectral/\\form?collection$\_$name=UVES}.
The reduced spectra are characterized by R = 42310 and S/N $\sim$ 19 - 183 at $\sim$ 921 nm.
The coordinates and the S/N of the data are listed in Table \ref{data}.

In order to minimize systematic errors, a homogeneous comparison between bulge and disk stars with atmospheric parameters estimated in a consistent way is required.
\cite{2014A&A...562A..71B} spectroscopically analyzed 714 F and G dwarf and subgiant stars in the nearby Galactic disk. Using kinematical criteria defined in \cite{2003A&A...410..527B}, they obtained for each star two relative probabilities: the thick-disk-to-thin-disk (TD/D) and the thick-disk-to-halo (TD/H) membership. They required a probability of TD/D$>2$ to classify a star as thick disk candidate, while TD/D$<0.5$ for a candidate thin disk star.
According to their kinematical criteria, their sample includes 387 stars with thin disk kinematics, 203 stars with thick disk kinematics, 36 stars with halo kinematics (TD/H$<1$) and 89 stars with kinematics between those of thin and thick disks.

We explored the UVES archive searching for high resolution (R>40000) and high signal-to-noise (S/N>100) spectra for the sample of 203 thick disk stars, in the wavelength range were S lines of Mults. 1, 6 and 8 are available. The reduced spectra of 23 thick disk stars were found and retrieved for the analysis. The data are characterized by R$\sim$42000-110000 and S/N$\sim$122-400 and they cover the wavelength range 665 nm - 1042 nm or 565 nm - 946 nm. In Table \ref{thick} are reported the names and coordinates of the thick disk stars from the Hipparcos catalog \citep{1997A&A...323L..49P}.

Among the 387 thin disk stars \citep{2014A&A...562A..71B}, 30 stars were observed during the UVES Paranal Observatory Project\footnote{https://www.eso.org/sci/observing/tools/uvespop/\\field$\_$stars$\_$uptonow.html} (UVES POP, \citealt{2003Msngr.114...10B}).
We obtained the UVES POP spectra, which cover the wavelength range 300-1000 nm and are characterized by R$\sim$80000 and S/N $\sim$300-500 in the V band.
The names and coordinates of UVES POP stars are reported in Table \ref{uvespop}.

\section{Stellar sample analysis}
The sample of stars considered in this work was already studied by \cite{2009A&A...499..737B, 2010A&A...521L..57B, 2010A&A...512A..41B, 2011A&A...533A.134B, 2013A&A...549A.147B, 2014A&A...562A..71B, 2017A&A...605A..89B, 2020A&A...634A.130B}.
They estimated atmospheric parameters, chemical abundances of 13 species (Fe, O, Na, Mg, Al, Si, Ca, Ti, Cr, Ni, Zn, Y, Ba), ages  and radial velocities of the targets.

\begin{figure*}
  \centering
	\includegraphics[trim= 1cm 5cm 1cm 5cm, clip, width=0.6\textwidth]{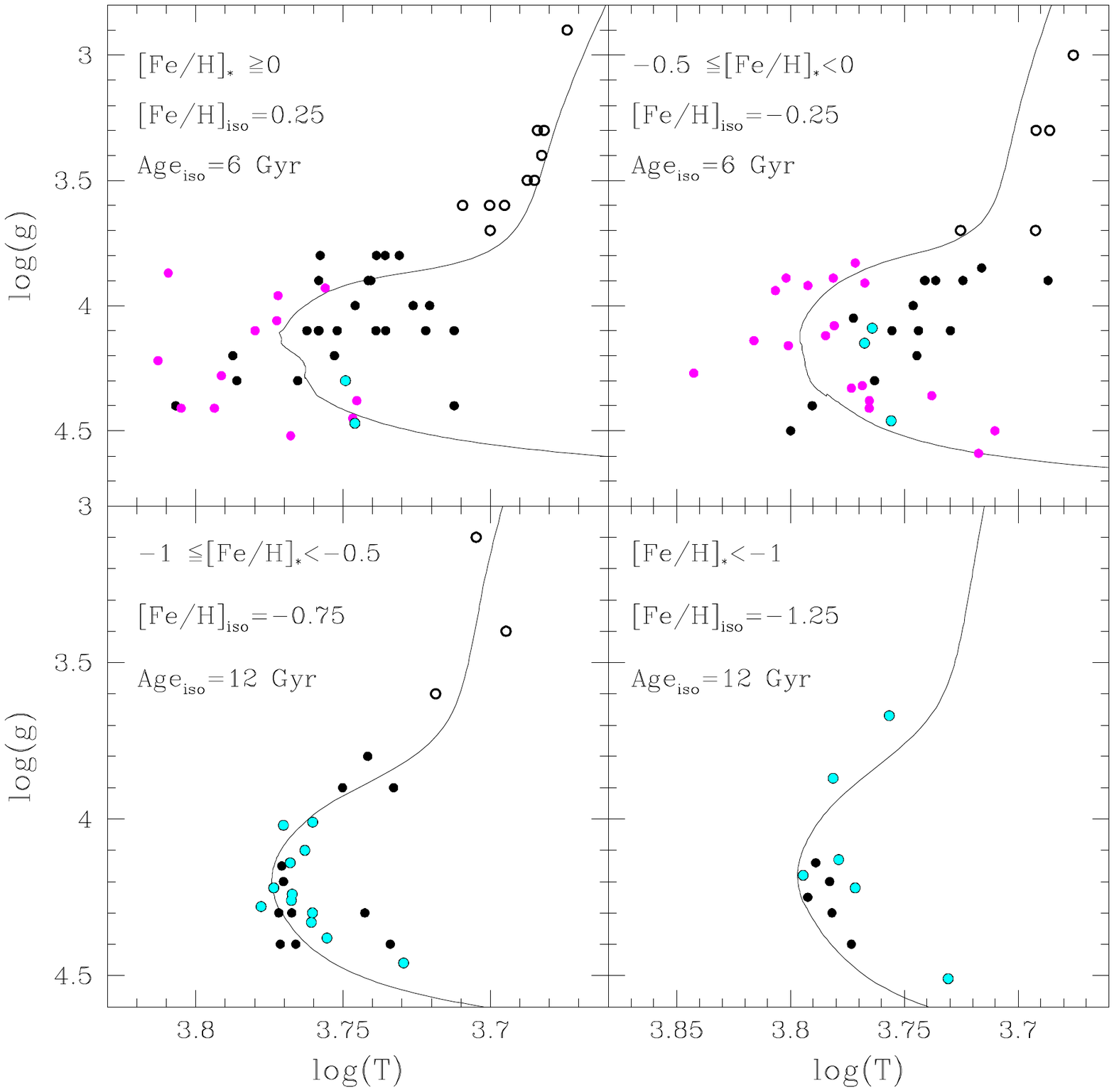}
	\caption{$\log(g)$ vs $\log(T_{eff})$ diagrams of the stellar sample considered in this work. The bulge, thick and thin disk stars are shown in black, cyan and magenta respectively. The black filled circle and open circles are bulge dwarfs and giants. In each panels are reported the metallicities and the ages adopted to obtain isochrones from the  Dartmouth Stellar Evolution Database \citep{2008ApJS..178...89D}.}
	\label{cmd}	
\end{figure*}

Using the atmospheric parameters measured by Bensby and collaborators, we created $\log(g)$ vs $\log(T_{eff})$ diagrams (Figure \ref{cmd}).
We divided the stellar sample in sub-samples with similar metallicities, dividing the metallicity range $-2<$[Fe/H]$<1$ in bins of 0.5 dex.
In Figure \ref{cmd} the bulge, thick and thin disk stars are shown in black, cyan and magenta, respectively.
In each panel are reported different isochrones from the Dartmouth Stellar Evolution Database \citep{2008ApJS..178...89D} that allow to guide the eye.
From Figure \ref{cmd} we identified dwarf and giant stars as points below and above $log(g)=3.7$ $g/cm^3$, respectively.
While the disk sample includes only dwarf stars, the bulge sample presents dwarf (filled circles) and sub-giant (open circles) stars.

\begin{figure}
  \centering
	\includegraphics[trim= 1.3cm 5.5cm 1cm 5cm, clip, width=0.45\textwidth]{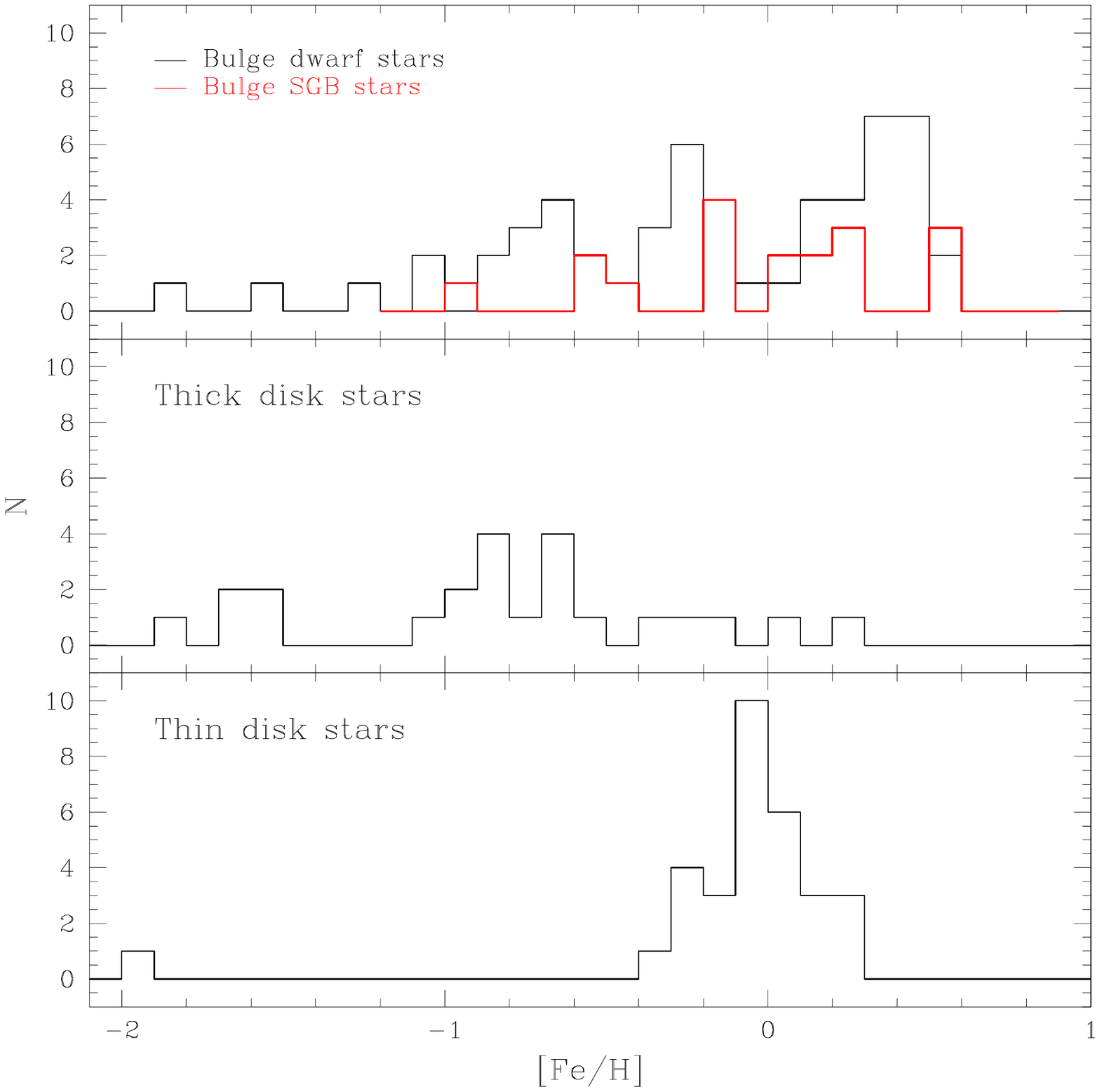}
	\caption{Metallicity distributions of bulge and disk stars. In the top panel are compared the metallicity distributions of bulge dwarfs (black) and giants (red).}
	\label{metdist}	
\end{figure}

Figure \ref{metdist} shows the metallicity distribution of bulge and disk stars.
In the top panel are compared the metallicity distribution of bulge dwarfs (black) and giants (red).

\begin{figure*}
  \centering
	\includegraphics[trim= 1cm 12cm 1.5cm 9.5cm, clip, width=0.9\textwidth]{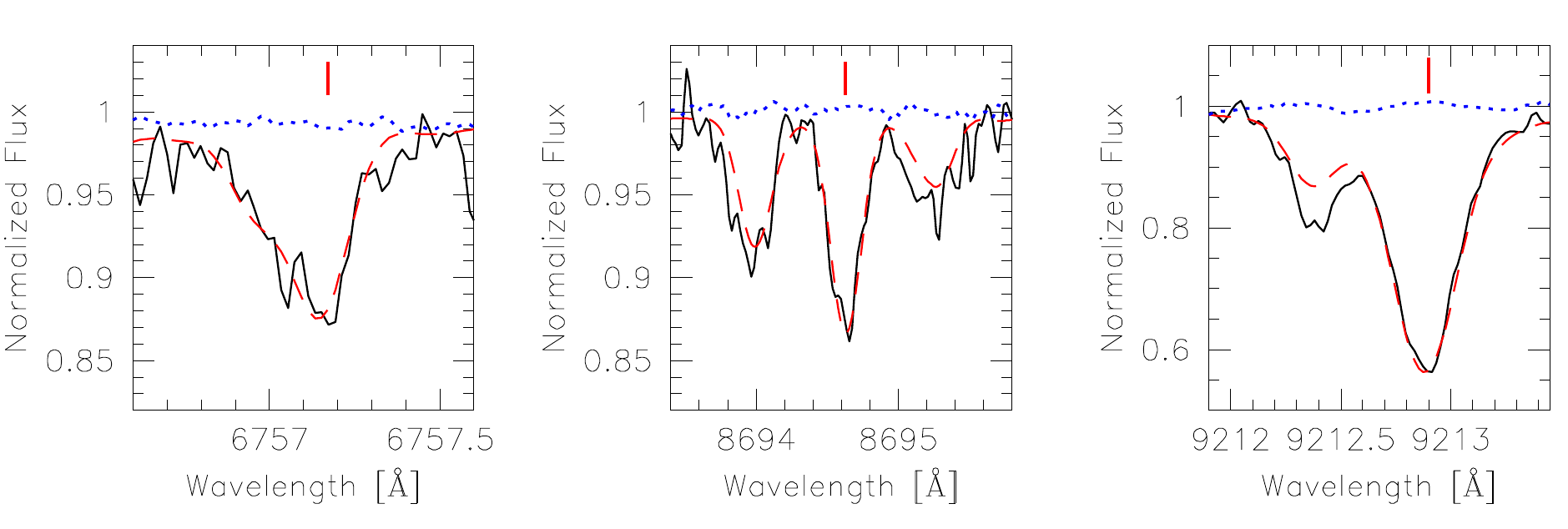}
	\caption{Observed spectrum of the bulge star OGLE-2014-BLG-1418S (black) superimposed by the best fit (dashed red) synthetic spectrum. The contribution due to the terrestrial atmosphere is also shown (dotted blue). In left, middle and right panels are reported S features (vertical red lines) of Mult. 8, 6 and 1, respectively.}
	\label{fit}	
\end{figure*}

\section{Chemical abundance analysis}
\subsection{Atmospheric parameters}
In this work, we decided to use atmospheric parameters and metallicities in the literature.

In Table \ref{data} are reported atmospheric parameters, metallicity and radial velocities from \cite{2017A&A...605A..89B} adopted for the 74 bulge stars. 

The atmospheric parameters and metallicities that we adopted from \cite{2014A&A...562A..71B} for the thick and thin disk stars are listed in Tables \ref{thick} and  \ref{uvespop}, respectively.

The considered bulge and disk stars represent a sample of targets characterized by atmospheric parameters estimated in a consistent way.

\subsection{Sulfur abundances}
The spectral range covered by the data allowed us to measure A(S) from lines of Mult. 1, 6 and 8 (Table \ref{lines}).

In order to identify the lines, we computed 
synthetic spectra with the code SYNTHE (\citealt{1993sssp.book.....K},\citeyear{2005MSAIS...8...14K}), using ATLAS9 $\alpha$-enhanced model atmospheres \citep{1993KurCD..13.....K} based on ODF by \cite{2003IAUS..210P.A20C} with the parameters in Table \ref{data}.
To assess the contamination due to telluric lines, we compared the observed spectra of bulge targets with those of B stars observed on the same night.
Similarly, we used UVES POP B stars spectra observed on the closer observation night of the disk stars. 
Thanks to the simple superimposition of these spectra, we evaluated the line suitability for the estimation of A(S). 
We rejected lines of Mult. 1 contaminated by telluric ones and Mult. 6 and 8 lines too weak to be analyzed. Moreover, we selected only stars with at least two S lines in their spectra. After this preliminary selection of S I lines, the sample of thick disk stars reduced to 21, while all the bulge and thin disk stars were considered for the analysis.

S abundances were derived from lines of Mult. 1, 6 and 8 by spectrosynthesis using our own code SALVADOR. This calculates a grid of synthetic spectra,
and finds the abundance that minimizes the $\chi^2$ with the observed spectrum.
Figure \ref{fit} shows the best fit obtained for S lines (red vertical lines) of Mult. 8 (left panel), Mult. 6 (middle panel) and Mult. 1 (right panel) for the target OGLE-2014-BLG-1418S.
In the cases where Mult. 1 lines were blended with telluric lines, we estimated A(S) from line equivalent width (EW) using the code GALA \citep{2013ascl.soft02011M}.
EWs were measured with the IRAF\footnote{IRAF is distributed by the National Optical Astronomy Observatory, which is operated by the Association of Universities for Research in Astronomy (AURA) under a cooperative agreement with the National Science Foundation.} task \textit{splot} and the contribution from the telluric and the S line were  taken into account
using the \textit{splot deblend} option.

To evaluate the A(S) uncertainty due to the atmospheric parameters errors, we assumed the mean errors on temperature, gravity, microturbulence velocity and metallicity estimated by \cite{2017A&A...605A..89B} with the method outlined in \cite{2010ApJ...709..447E}.
A $\sigma_T \sim$94 K, $\sigma_{log(g)} \sim$0.15 and $\sigma_{v_m} \sim$0.17 km s$^{-1}$ lead to an A(S) uncertainty of 0.1 dex, 0.04 dex and 0.03 dex, respectively. 
The A(S) error due to a variation of 0.13 dex in metallicity is 0.03 dex.
 
In order to consider the deviation from LTE, we applied Non-LTE (NLTE) corrections to lines of Mult. 1 and 6, according to \cite{2005PASJ...57..751T}. 
Indeed, the NLTE effects are negligible for lines of Mult. 8, while they are small and moderate for Mult. 6 and 1. Moreover, the NLTE correction increases as the effective temperature increases and the surface gravity decreases \citep{2005PASJ...57..751T, 2020MNRAS.496.2462K}. 
As expected, we found that Mult. 6 lines are less affected by LTE deviations than those of Mult. 1. Indeed we obtained NLTE corrections of about $-0.06<\Delta<-0.01$ and $-0.26<\Delta<-0.1$ for features of Mult. 6 and 1, respectively.
Overall, the mean NLTE corrections are $<\Delta> \sim-0.02$ dex for Mult. 6 and $<\Delta> \sim-0.15$ dex for Mult. 1. 
Finally, we calculated the [S/Fe] adopting the solar value A(S)$_{\odot}$=7.16 \citep{2011SoPh..268..255C}. We report in Table\,\ref{data} the measured A(S)$_{\rm NLTE}$, the standard deviation, the number of lines used and the [S/Fe] ratio.

\begin{figure}
  \centering
	\includegraphics[trim= 1cm 5.5cm 0.8cm 4.5cm, clip, width=0.5\textwidth]{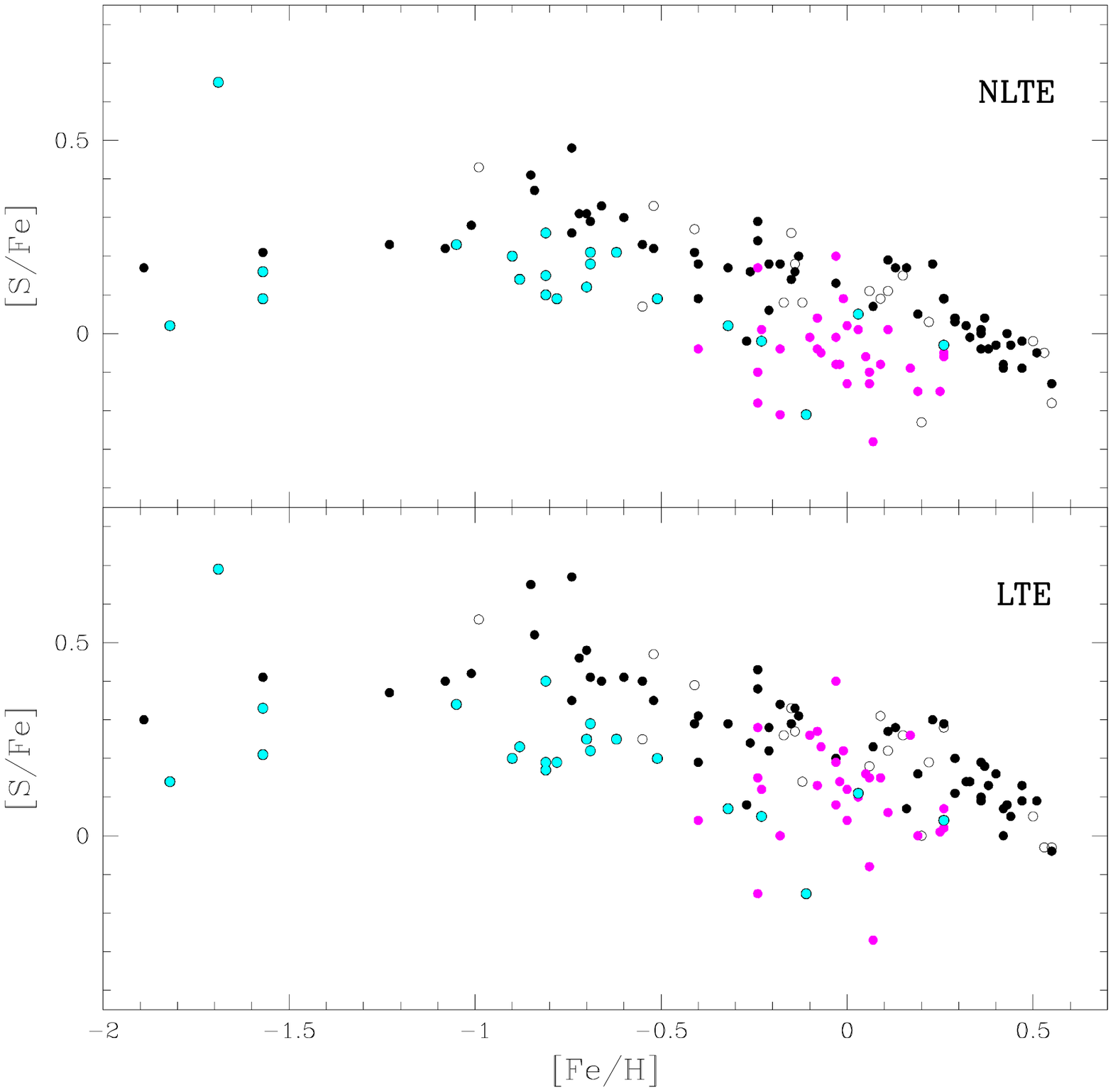}
	\caption{[S/Fe] vs [Fe/H] diagram before (bottom panel) and after (top panel) NLTE corrections applied to the A(S) obtained in this work for bulge (black), thick disk (cyan) and UVES POP thin disk (magenta) stars. The bulge dwarfs and giants are shown as filled and open circles, respectively.}
	\label{knee}	
\end{figure}

From Fig. \ref{knee} it is possible to appreciate the differences in the [S/Fe] vs [Fe/H] diagram due to NLTE corrections. The bulge dwarf and giant stars are shown as black points and black open circles. The cyan and magenta points are thick and thin disk stars. Despite NLTE corrections lead to a difference in [S/Fe] values, the relative distribution in the [S/Fe] vs [Fe/H] diagram does not change. As mentioned before, the NLTE correction depends on the temperature, surface gravity and metallicity. In spite of this we find similar mean NLTE corrections ($\Delta \sim -0.13$) for bulge dwarfs and giants stars. In comparison to bulge stars, the thin disk ones show a larger mean NLTE correction, $\Delta \sim -0.15$. This effect is probably due to the temperature since thin disk stars are dwarfs, thus younger and hotter stars.

\begin{figure}
  \centering
	\includegraphics[trim= 0.8cm 6cm 0.8cm 5cm, clip, width=0.5\textwidth]{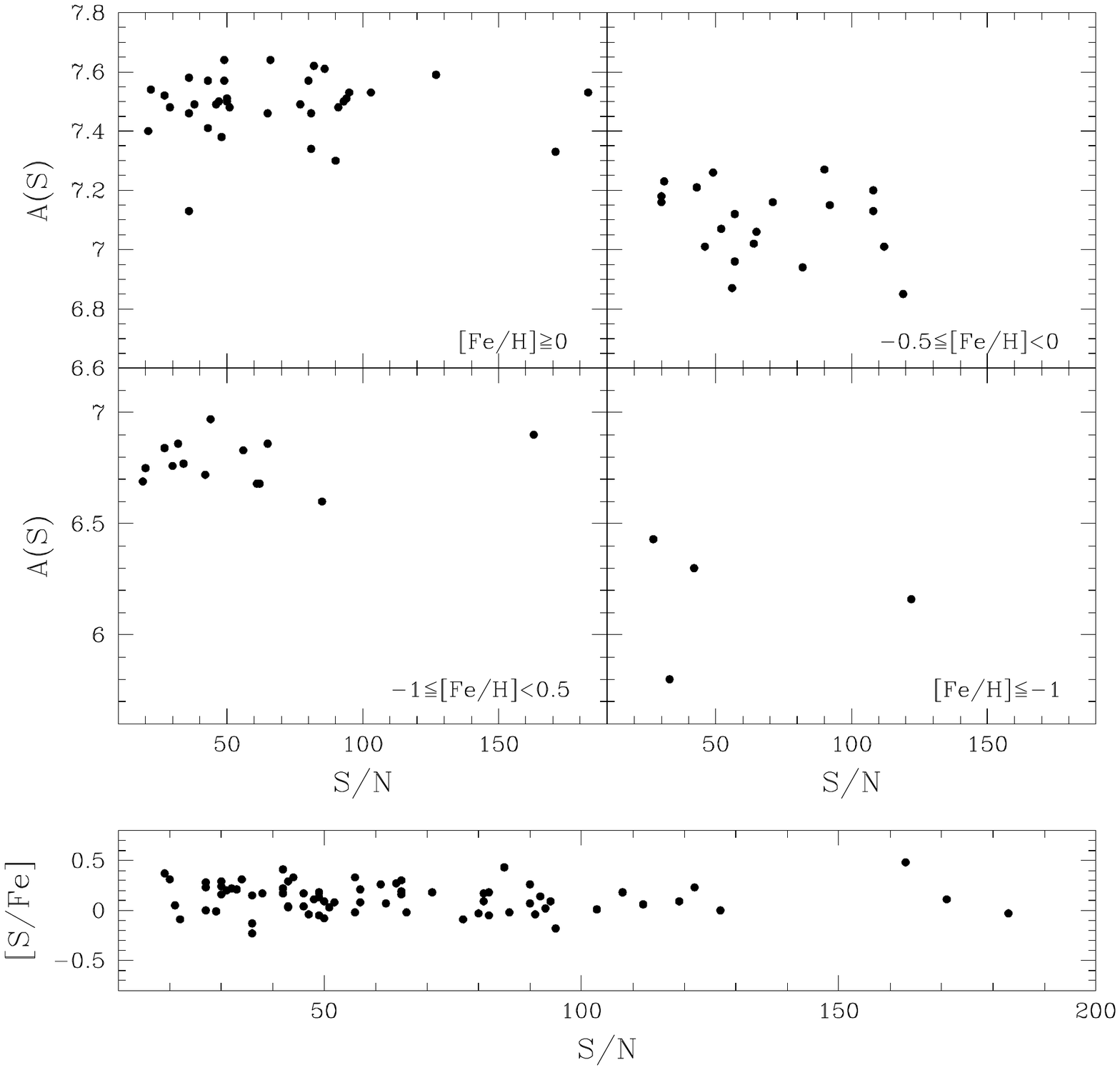}
	\caption{A(S) vs S/N and [S/Fe] vs S/N diagrams.}
	\label{snr}	
\end{figure}

Since our bulge sample is characterized by a large range in S/N, we investigated any dependence on this value. We divided the bulge stars in sub-samples with similar metallicity considering [Fe/H] bins of 0.5 dex to compare the A(S) with the S/N. As shown in Figure \ref{snr}, the scatter in A(S) increases with the decreasing of S/N and no trends are found. Similarly, we do not found dependencies between [S/Fe] and the S/N. Finally, we want to underline that the A(S) of stars with S/N around 20 were measured from Mult. 1 lines, which are the strongest in stellar spectra.

To measure the A(S)$_{\rm NLTE}$ and the [S/Fe] of disk stars, we followed the same procedure above described for bulge stars. The results obtained for thick and thin disk stars are reported in Table \ref{thick} and \ref{uvespop}, respectively. The atmospheric parameters and the metallicities are from \cite{2014A&A...562A..71B}. Following \cite{2014A&A...562A..71B}, we adopted $\sigma_T \sim$100 K, $\sigma_{log(g)}\sim$0.1 dex, $\sigma_{v_m}\sim$0.1 kms$^{-1}$ and $\sigma_{[Fe/H]}\sim$0.1 dex to evaluate the S abundance uncertainty of disk stars due to atmospheric parameters errors, obtaining variations of 0.03 dex, 0.01 dex, 0.02 dex and 0.01 dex, respectively.

\section{Discussion}
\subsection{Sulfur behavior in the Galactic bulge}
Several works confirm that S behaves like an $\alpha$-element in the disk and halo of the MW (\citealt{1987A&A...176..294F}, \citealt{2004A&A...415..559R},  \citealt{2007A&A...470..699C}, \citealt{2011A&A...528A...9S}, \citealt{2016PASJ...68...81T}).
It is observed that from a constant value of [S/Fe]$\sim$+0.4 at low metallicity, [S/Fe] decreases with increasing  [Fe/H] and  it reaches [S/Fe]$\sim$0 at solar metallicities (\citealt{2007A&A...469..319N}, \citealt{2017A&A...604A.128D}).
In contrast, other works found different S trends in the low-metallicity regime.
The [S/Fe] value obtained by \cite{2001ApJ...557L..43I} constantly increases as metallicity decreases, up to  $\sim$0.7-0.8 dex at $-2.3<$ [Fe/H] $<-1.9$.  
\cite{2005A&A...441..533C} found a bimodal behavior of [S/Fe] (both stars with [S/Fe]$\sim+$0.4 and higher values) at low metallicity.

In Figure \ref{knee} are compared the [S/Fe] versus [Fe/H] diagrams obtained before (bottom panel) and after (top panel) NLTE corrections for bulge stars (black).
Both diagrams show that [S/Fe] decreases with the increase of  [Fe/H] 
and that the bulge remains S-enhanced at solar metallicity.
In the metal-poor regime, the low number of stars does not allow us 
to draw firm conclusions about the presence of a plateau.
Considering the trend shown in Figure \ref{knee}, we can guess the presence of a plateau 
at the value [S/Fe]$_{\rm NLTE}$ $\sim+$0.2 for  [Fe/H]$<-1$. 
On this basis, we claim that S behaves like an $\alpha$-element in the bulge.
However, a larger number of metal-poor stars is required to 
robustly confirm or reject this claim.

Recently, \cite{2020arXiv200905063G} analyzed abundance ratio trends of sixteen species in Galactic bulge and thick disk RGB stars from the Apache Point Observatory Galactic Evolution Experiment (APOGEE, \citealt{2017AJ....154...28B}, \citealt{2017AJ....154...94M}) spectra.
Among the different elements, the investigation of S in bulge stars was performed for the first time.
They derived A(S) for $\sim$11000 targets in the metallicity range $-1.5<$[Fe/H]$<1$ using \ion{S}{i} lines at 15478.5 $\AA$ and 16576.6 $\AA$, which are affected by small NLTE deviations \citep{2018AJ....156..126J, 2020AJ....160..120J, 2020MNRAS.496.2462K}.

\begin{figure}
  \centering
\includegraphics[trim= 1cm 5.5cm 0.8cm 5cm, clip, width=0.5\textwidth]{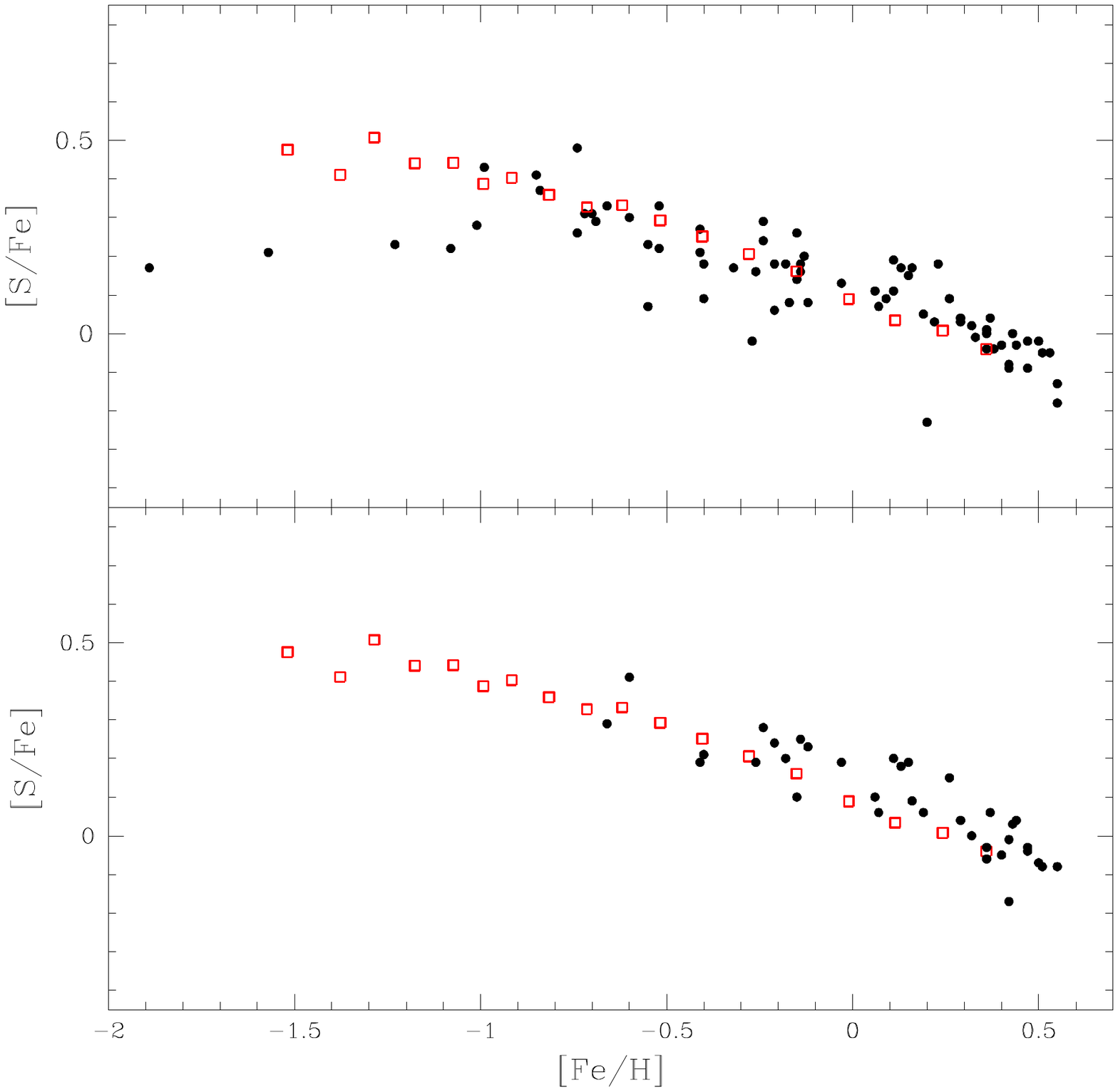}
\caption{[S/Fe] versus [Fe/H] diagram. The results obtained in this work for bulge stars (black) are compared with the mean trend (open red squares) of \cite{2020arXiv200905063G}. [S/Fe]$_{\rm NLTE}$ values of bulge stars were calculated considering features of Mults. 1, 6 and 8 and features of Mults. 6 and 8 in the upper and bottom panel, respectively.}
\label{SFe}	
\end{figure}

In Figure \ref{SFe}, we compare our [S/Fe]$_{\rm NLTE}$ (black) calculated considering features of Mults. 1, 6 and 8 (upper panel) and Mults. 6 and 8 only (bottom panel) with the median trend (open red squares) obtained by \cite{2020arXiv200905063G}.
In the bottom panel, our sample of bulge stars decreases to 37 due to the exclusion of Mult. 1, the strongest of the measured S lines and the most affected by NLTE deviations. However, we found similar S behavior in bulge stars at [Fe/H]$>-0.2$ when we consider Mult. 1, 6 and 8 (upper panel) corrected for NLTE effects and features less affected by NLTE deviation (bottom panel) to calculate S abundances.

Our results and those of \cite{2020arXiv200905063G} are similar in the metallicity range $-1<$[Fe/H]$<-0.1$, while 
on average our [S/Fe] ratios are higher in the range
[Fe/H]$>0.1$ by about 0.04 $\pm$ 0.01 dex.

\begin{figure}
  \centering
	\includegraphics[trim= 0.8cm 5.5cm 0.8cm 5cm, clip, width=0.5\textwidth]{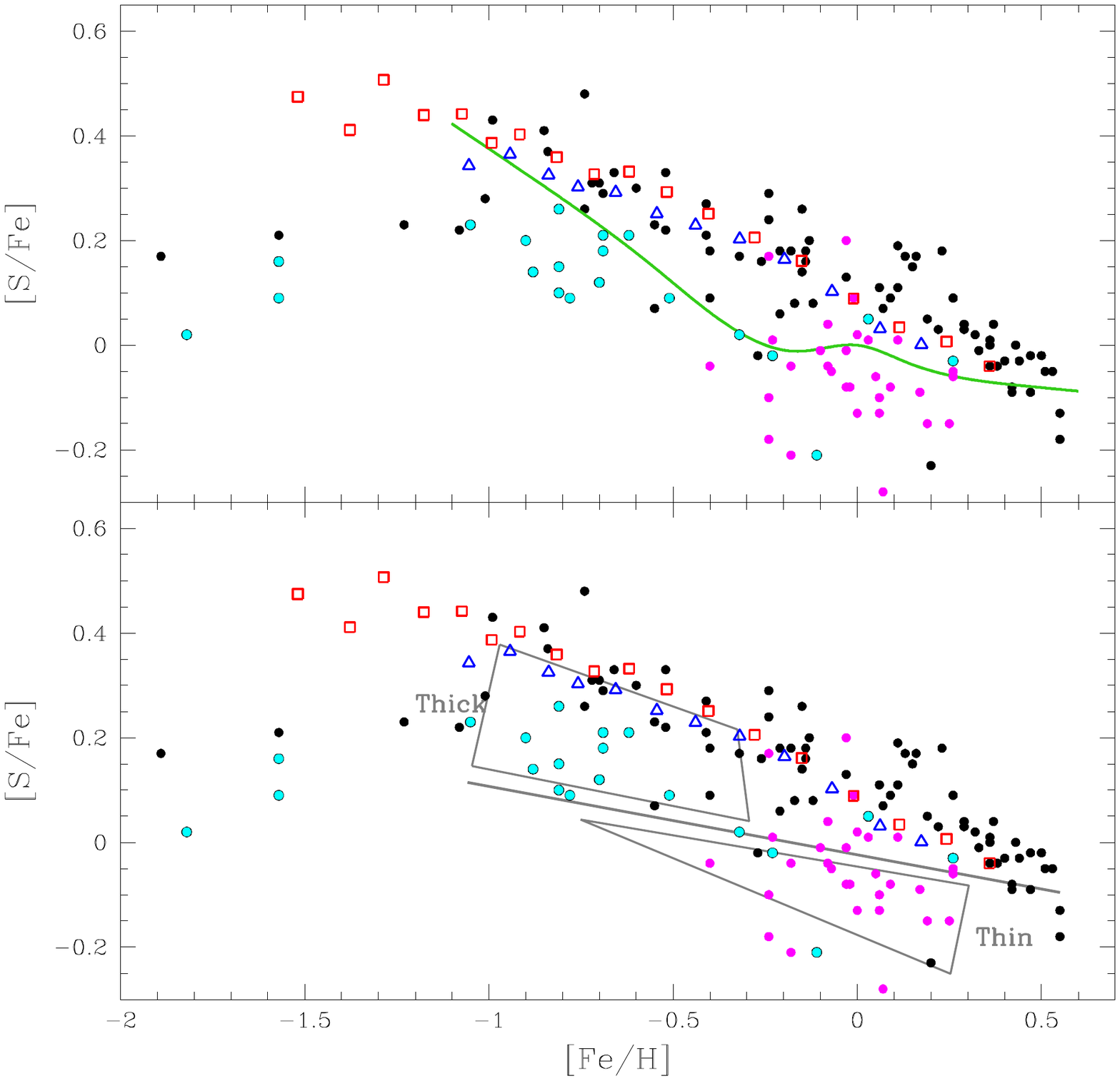}
	\caption{[S/Fe] versus [Fe/H] diagrams. The [S/Fe]$_{\rm NLTE}$ measured in this work for bulge, thick and thin dwarf stars are shown in black, cyan and magenta points, respectively. The open red squares and open blue triangles are the median trends for bulge and thick disk stars studied by \cite{2020arXiv200905063G}. The mean trend obtained by \cite{2017A&A...604A.128D} for Galactic disk stars is reported as a green line. In the bottom panel are reported the thick and thin disk areas (gray) defined by \cite{2021arXiv210201961P}, and the line separating between them (gray continuous line).}
	\label{litNLTE}	
\end{figure}

\subsection{Comparison with Sulfur abundances in the disk}
In order to perform a homogeneous analysis of the bulge and the disk, a comparison sample of disk stars with stellar parameters derived in similar way is required.
Indeed, differences in the choice made to estimate the stellar atmospheric parameters, particularly in the effective temperature scale, may lead to systematic differences in the derived abundances among different studies.

With the aim of overcoming this problem, we decided to derive the S abundance of a kinematically selected sample of thick and thin disk stars whose atmospheric parameters were determined consistently with our bulge stars.

We considered 21 thick disk stars and 30 thin disk ones previously analyzed by \cite{2014A&A...562A..71B}.
The thick disk stars are located between 14.8 pc and 109.9 pc from the Sun and within 2410 pc from the Galactic plane. Whereas, the thin disk stars have heliocentric distance between 7.5 pc and 72 pc and a maximum distance from the Galactic plane of 680 pc.

The LTE (lower panel) and NLTE (upper panel) [S/Fe] values obtained from Mults. 1, 6 and 8 for the thick (cyan) and thin (magenta) disk stars are reported in Figure \ref{knee}. 
The comparison in Figure \ref{knee} shows that the bulge is S-rich with respect to both the thick and thin disks.

Using the Gaia ESO Survey data \citep{2012Msngr.147...25G}, \cite{2017A&A...604A.128D} measured S abundances of F and G-type stars in the solar neighborhood.
Their dwarf stars sample is located within $\leq 1.5$ kpc from the Sun at Galactic latitude $\langle\vert b \vert \rangle\approx 30^{\circ}$, while giants cover a larger range in distance, $0<{\rm D/kpc}< 16$, at $\langle\vert b \vert \rangle\approx 9^{\circ}$.
They found that their sub sample of dwarfs is dominated by thick disk stars at [Fe/H]$<-0.5$, while giant stars mainly belong to the thin disk and dominate the sample at higher metallicities.

Recently, \cite{2021arXiv210201961P} investigated the evolution of S in the Milky Way using FGK-type stars spectra provided by 
the Arch\'eologie avec Matisse Bas\'ee sur les aRchives de l 'ESO (AMBRE) 
Project \citep{2013Msngr.153...18D}.
Their sample of stars is located within 200 pc from the Sun and within a distance smaller than 400 pc from the Galactic plane.
They identified two disk components characterized by different age, kinematics and sulfur abundances.
In order to distinguish the thin and thick disk, they defined a separation line
in the [Fe/H] vs. [S/Fe] plane.
They associated the metal-poor, S-rich stars above the separation
line to the thick disk.
The metal-rich, S-poor stars below the separation line 
were associated to the thin disk. 
According to their work, the thick disk component has low rotational velocities and is older than the thin disk 
The thick disk is also  S-rich with respect to the thin disk in the metallicity range $-1\leq$[Fe/H]$\leq-0.5$. 
The thin disk stars, instead, 
have rotational velocities close to the solar one and S abundances that slowly decrease from [Fe/H]$\sim-0.5$\,dex up to +0.5\,dex.

\cite{2020arXiv200905063G} defined the Galactic thick disk as stars with Galactocentric radius $5$ kpc $<\rm{R_{GC}}<11$ kpc and mid-plane distance $\vert Z \vert<2$ kpc. Comparing the S evolution in the bulge and the thick disk they found similar S median trends.

In Figure \ref{litNLTE} we compare our results with those mentioned above in the [S/Fe] versus [Fe/H] diagram.
The bulge, thick and thin disk stars analyzed in this work are reported as black, cyan and magenta points, respectively.
The green line in the upper panel is the mean trend of disk stars analyzed by \cite{2017A&A...604A.128D} and is dominated by thick disk stars at [Fe/H]$<-0.5$ and by thin disk ones at higher metallicities.
The separation line (gray) and the thick and thin disk areas (gray) defined by     \cite{2021arXiv210201961P} are shown in the bottom panel.
In the upper panel the bulge and thick disk stars studied by \cite{2020arXiv200905063G} are reported as red open squares and blue open triangle symbols.

In the metallicity range -1<[Fe/H]<-0.5, the [S/Fe] values obtained in this work for thick disk stars lie in the thick disk area defined by \cite{2021arXiv210201961P}. On the other hand, we found that our thick disk sample is 0.09 $\pm$ 0.05 dex and 0.14 $\pm$ 0.01 dex less S-enriched with respect to the thick disk sample of \cite{2017A&A...604A.128D} and \cite{2020arXiv200905063G}, respectively.

Despite the low number of thick disk stars at low and super-solar [Fe/H], we found that the bulge is characterized by a higher S content than the thick disk in the metallicity range -1<[Fe/H]<-0.5. This outcome is at odds with
the result of \cite{2020arXiv200905063G}, who found similar S trends for bulge (red open squares) and thick disk (blue open triangles) stars.

Finally, the thin disk stars are slightly less S-rich by 0.04 $\pm$ 0.02 dex than those studied by \citealt{2017A&A...604A.128D}, while they are similar to the thin disk population of \cite{2021arXiv210201961P}. We observe a substantial  agreement between the thin and thick disk stars and the literature and we conclude that the bulge is S-enriched with respect to the thin and thick  disk.

\subsection{Comparison between $\alpha$-elements in the bulge and the disk}
From Figure \ref{litNLTE}, it is clear that the trend in the [S/Fe] vs [Fe/H] plane is different for bulge and thin disk stars.
The situation is, on the other hand, less clear when considering the thick disk.
While we obtain different S content for bulge and thick disk stars,  \cite{2020arXiv200905063G} found a substantial agreement between their bulge and thick disk stars.
Notice that the A(S) of the disk stars we analyzed are comparable with those of \cite{2017A&A...604A.128D} and \cite{2021arXiv210201961P}.

The evidence of different [$\alpha$/Fe] abundance trends between the bulge and the thin and thick disk was found by other studies, considering both dwarf (\citealt{2017A&A...605A..89B}) and RGB (\citealt{2005ApJ...634.1293R}, \citealt{2008MmSAI..79..503Z}) stars.
According to \cite{2006ApJ...651..491C}, the [O/Fe] and [Ti/Fe] trends of K and M RGB stars in the bulge fall above those of the thin and thick disk.
\cite{2007ApJ...661.1152F} derived detailed abundances of O, Na, Mg, Al, Si, Ca and Ti from high-resolution HIRES spectra of 27 RGB stars observed toward the Galactic bulge in the Baade's window.
They found that in the metallicity range $-1.5<$[Fe/H]$<0.5$ all the mean trends of $\alpha$-elements in the bulge lie $\sim$0.2 dex higher than that of the thin disk.
\cite{2014AJ....148...67J} analyzed the high resolution and high S/N FLAMES/GIRAFFE spectra of 156 RGB stars to estimate Mg, Si, Ca and Ti abundances.
Overall, they found that the bulge exhibits higher [$\alpha$/Fe] ratios than the local thick disk at [Fe/H]$\leqslant-0.5$.
From APOGEE spectra of stars located in the Baade's Window, \cite{2017A&A...600A..14S} obtained a median trend for bulge stars more enhanced than the thick disk one in the [Mg/Fe] vs [Fe/H] diagram.

However, there is no universal agreement in the literature that the [$\alpha$/Fe] trends of bulge and thick disk are different.
According to \cite{2010A&A...513A..35A}, the bulge and thick disk follow the same [$\alpha$/Fe] trend in the metallicity range $-1.5<$[Fe/H]$<-0.3$.
This result is in good agreement with \cite{2011A&A...534A..80H}, who found similar [Mg/Fe] trends in bulge and thick disk at [Fe/H]$<-0.4$.
On the other hand, the bulge shows higher [Mg/Fe] values than thick and thin disk ones at $-0.4<$[Fe/H]$<-0.1$, while similar [Mg/Fe] ratios between bulge and local thin disk are found at [Fe/H]$>-0.1$ (\citealt{2011A&A...534A..80H}).
\cite{2011A&A...530A..54G, 2015A&A...584A..46G} 
found that bulge and thick disk have 
similar abundances of $\alpha$-elements  at low metallicity. 
Both are enhanced with respect to the thin disk.
At solar metallicities the bulge presents [$\alpha$/Fe] ratios similar to those of the thin disk.

Thanks to the Gaia-ESO survey, \cite{2017A&A...601A.140R} performed a homogeneous comparison between the bulge, thin and thick disk sequences in the [Mg/Fe] vs. [Fe/H] diagram. 
In this context, the bulge and thick disk stars show similar [Mg/Fe] ratio levels over the whole metallicity range spanned in common, while the bulge results Mg-enriched with respect to the thin disk at [Fe/H]$<$0.1 dex.
\cite{2017A&A...598A.101J} performed a homogeneous analysis on $\alpha$-elements in 46 and 291 K-type giants located in the bulge and local thick disk, respectively, founding no different trends for these components of the MW.

\begin{figure}
  \centering
	\includegraphics[trim= 0.8cm 5.5cm 0cm 5cm, clip, width=0.5\textwidth]{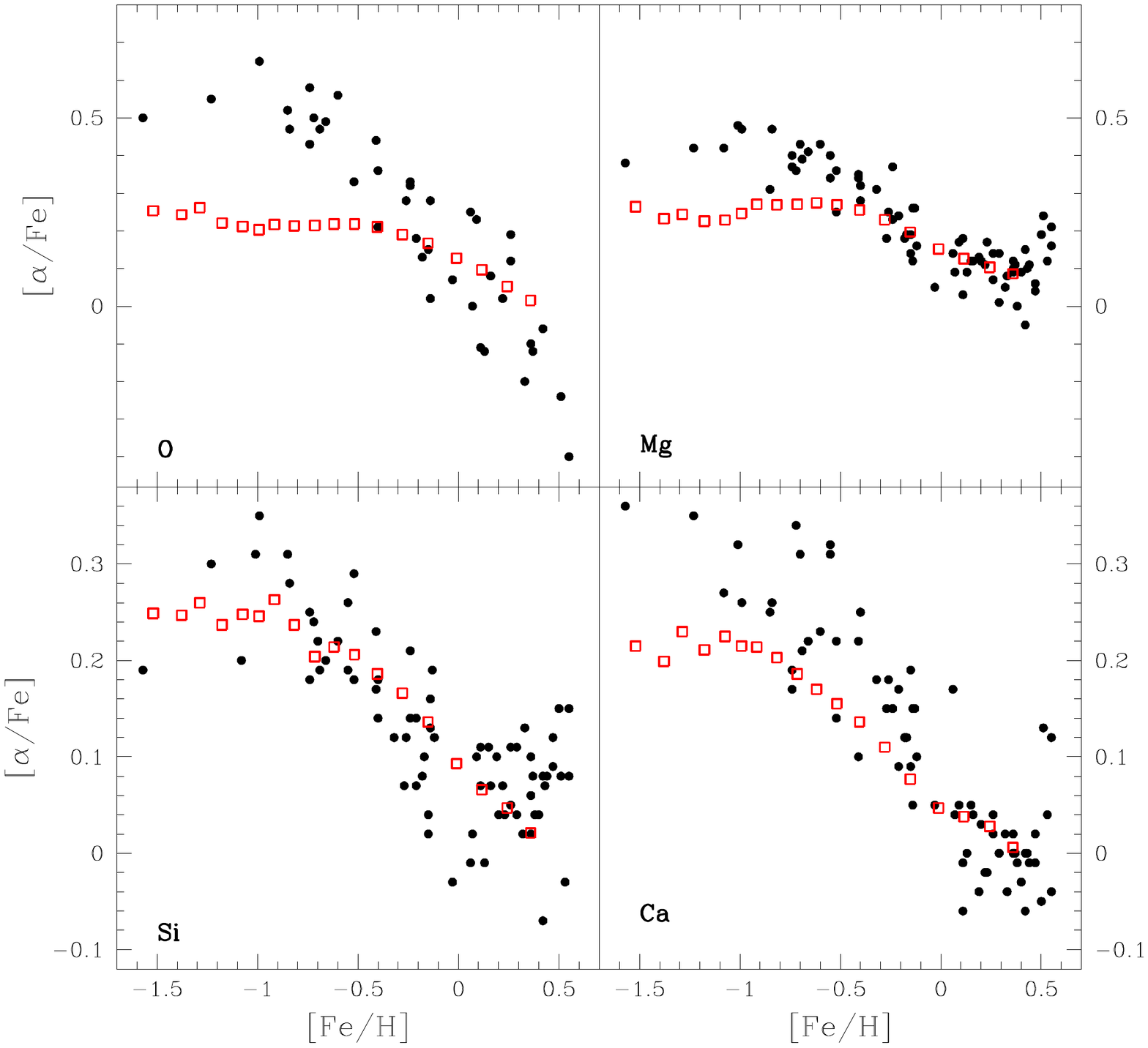}
	\caption{[$\alpha$/Fe] vs [Fe/H] diagrams. The O, Mg, Si and Ca abundances measured by \cite{2017A&A...605A..89B} for the bulge stars (black) analyzed in this work are compared with those by \cite{2020arXiv200905063G} (open red squares).}
	\label{alpha}	
\end{figure}

Finally, we note that, while for the sample of
microlensed bulge stars, we find a mean [S/Fe] vs [Fe/H] trend similar to \citet{2020arXiv200905063G}, the same is not true 
for the other $\alpha$-elements of the same stars. 
A significantly different trend can be appreciated at low metallicity 
between \citet{2020arXiv200905063G}
and \citealt{2017A&A...605A..89B} (see O, Mg and Ca in Figure \ref{alpha}).
A detailed analysis of the differences between these two studies is beyond the scope of the present contribution. Most likely this is rooted in the different spectral ranges, atomic/molecular data and the different analysis methods employed.
In Figures \ref{SFe}, \ref{litNLTE} and \ref{alpha} we show for comparison the mean trends of various [$\alpha$/Fe] abundance ratios obtained from APOGEE data. Notice, though, that the star to star scatter can be quite significant (see, for instance, Figure 3 in \citet{2020arXiv200905063G} and Figure 2 in \cite{2020MNRAS.496.2462K}.

\begin{figure}
  \centering
	\includegraphics[trim= 0.8cm 5.5cm 0cm 5cm, clip, width=0.5\textwidth]{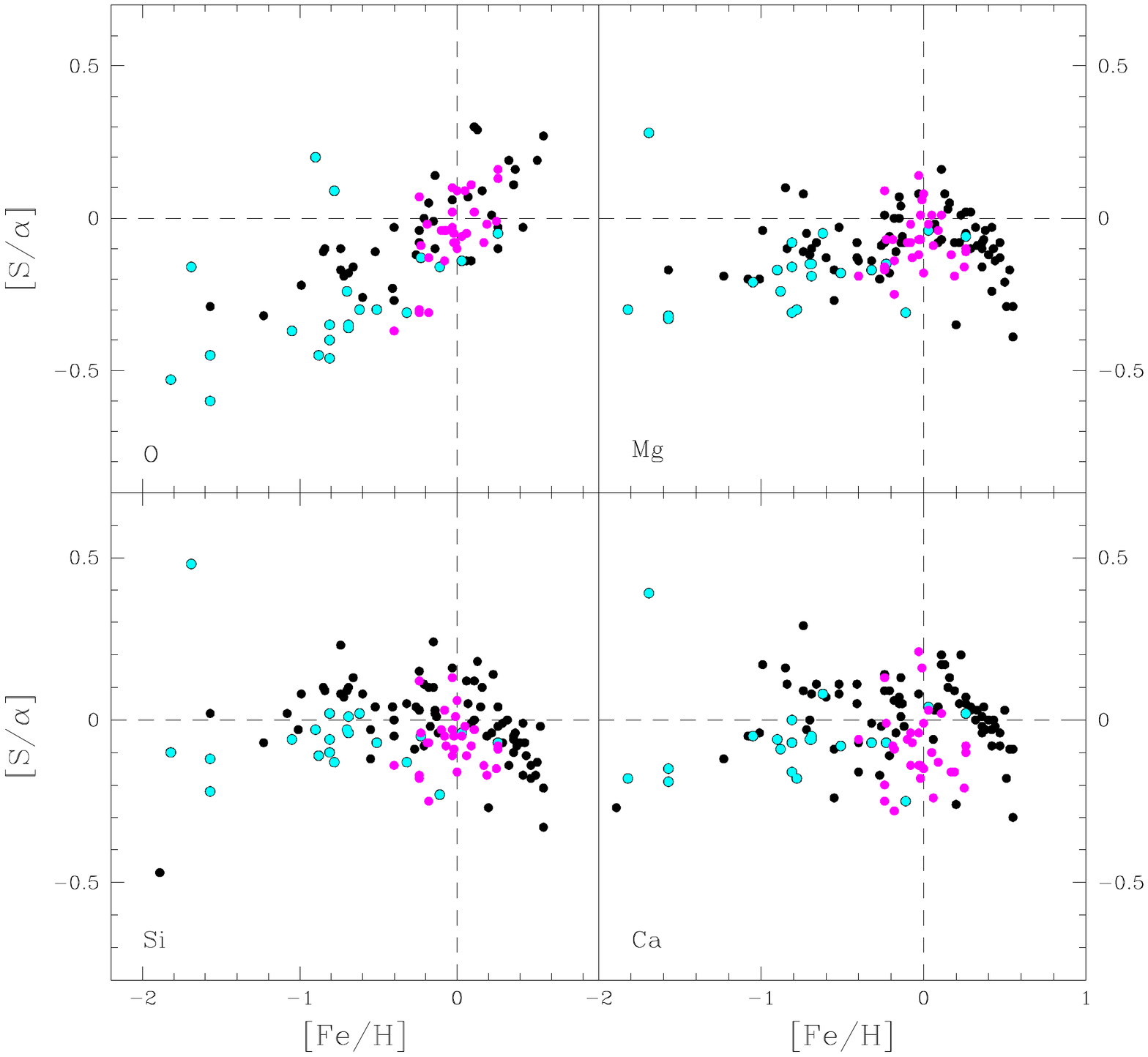}
	\caption{[S/$\alpha$] vs [Fe/H] diagrams of bulge (black), thick (cyan) and thin (magenta) disk stars. The $\alpha$-elements abundances and metallicities are from \cite{2014A&A...562A..71B, 2017A&A...605A..89B}. }
	\label{S_alpha}	
\end{figure}

In order to investigate the different $\alpha$-elements behavior, we created [S/$\alpha$] vs [Fe/H] diagrams. In Figure \ref{S_alpha} are compared the bulge (black), thick (cyan) and thin (magenta) disks trends. We found flat trends for the ratios of S over Mg, Si and Ca for bulge and disks samples, meaning that these elements are produced with no nuclosynthesis differences. On the other hand, more S is produced with respect to O as the metallicity increases.
One should however be aware that the O abundances of \citet{2014A&A...562A..71B}
were determined from the \ion{O}{i} permitted triplet lines
that are strongly affected both by departures from LTE 
and granulation effects \citep[see e.g.][and references therein]{2015A&A...583A..57S,2019A&A...630A.104A}, the effects
become larger at lower metallicity.
\citet{2014A&A...562A..71B} adopted the empirical correction of
\citet{2004A&A...415..155B} to correct their LTE oxygen abundances, one should
however be aware that the sample of stars that define this correction has
[Fe/H] $> -0.6$ thus beyond this limit the correction is extrapolating.

\section{Summary and conclusions}
This work investigates the behavior of Sulfur in the Galactic bulge.
We used high resolution high signal to noise spectra of 74 dwarf and sub-giant stars, collected with UVES during microlensing events and investigated by \cite{2017A&A...605A..89B}.
A reference sample of 21 and 30 F and G thick and thin disk stars located in the solar neighborhood was also analyzed.
For all samples, we used the atmospheric parameters estimated by Bensby et al. (\citeyear{2014A&A...562A..71B}, \citeyear{2017A&A...605A..89B}, \citeyear{2020A&A...634A.130B}) to measure A(S) from Mults. 1, 6 and 8.

Sulfur behaves like the others $\alpha$-elements in the bulge.
While we confirm the trend of [S/Fe] with [Fe/H] found by \cite{2020arXiv200905063G} in the bulge below metallicities [Fe/H]$< -$0.1, at higher metallicities our [S/Fe] measurements are slightly, but systematically, higher.

In order to compare bulge and disk stars whose atmospheric parameters were determined consistently, we measured A(S) of 21 and 30 thick and thin disk stars previously studied by \cite{2014A&A...562A..71B}.
In the metallicity range $-1<$[Fe/H]$<-0.5$, our measurements in thick disk stars are in agreement with \cite{2021arXiv210201961P} and they  agree within errors with \cite{2017A&A...604A.128D}. On the other hand, our sample of thick disk stars is S-poor with respect to \cite{2020arXiv200905063G}. Overall, we found that the bulge has a higher S content than the thick disk. This outcome contradicts the results of \cite{2020arXiv200905063G} but it is in agreement with what found for other $\alpha$-elements from other works \citep{2005ApJ...634.1293R, 2006ApJ...651..491C, 2007ApJ...661.1152F, 2008MmSAI..79..503Z, 2014AJ....148...67J, 2017A&A...605A..89B, 2017A&A...600A..14S}.

The [S/Fe] values obtained for thin disk stars are 
comparable with those by \cite{2017A&A...604A.128D} and \cite{2021arXiv210201961P}.
Our measurements imply that the bulge is S enriched with respect to the thin disk at [Fe/H]$>-0.4$, in agreement with previous works \citep{2007ApJ...661.1152F, 2014AJ....148...67J, 2011A&A...530A..54G, 2015A&A...584A..46G, 2017A&A...600A..14S, 2017A&A...601A.140R}.

In conclusion, our sulfur abundances support a scenario where the bulge and the disk experienced different chemical enrichment and evolution.
In particular, the S enhancement of bulge stars suggests that it formed more rapidly than the disk.

\longtab{
\centering
\small
\begin{landscape}
\begin{longtable}{p{0.3cm}p{3.4cm}p{1.8cm}p{1.8cm}p{0.6cm}p{1.7cm}p{1.7cm}p{1.7cm}p{1.7cm}p{1cm}p{0.4cm}p{1.7cm}p{1.7cm}}
\caption{Summary of bulge stars data. The target names, their coordinates and the S/N are listed in columns 2-5. The atmospheric parameters and radial velocities estimated by \citealt{2017A&A...605A..89B} follow in columns 7-11. The number of Sulfur line (N) used to estimate the NLTE Sulfur abundances obtained in this work are reported in the last two columns. The uncertainties on A(S) are the standard deviation.}\\
\label{data} \\
	\hline\hline
	&Object&RAJ2000&DECJ2000&S/N&T$_{eff}$&log(g)&$\xi$&[Fe/H]&RV&N&A(S)$_{\rm NLTE}$&[S/Fe]\\
	&&[hh:mm:ss]&[dd:mm:ss]&&[K]&[cgs]&[kms$^{-1}$]&&[kms$^{-1}$]&&&\\
	\hline
	\endfirsthead
	
	\bfseries \tablename \thetable{.continued}\\
	\hline
	&Object&RAJ2000&DECJ2000&S/N&T$_{eff}$&log(g)&$\xi$&[Fe/H]&RV&N&A(S)$_{\rm NLTE}$&[S/Fe]\\
	&&[hh:mm:ss]&[dd:mm:ss]&&[K]&[cgs]&[kms$^{-1}$]&&[kms$^{-1}$]&&&\\
	\hline
	\endhead
		
	\hline \multicolumn{13}{r}{Continue on next page} \\
	\endfoot
	
	\hline\hline
	\endlastfoot
1   & OGLE-2009-BLG-076S & 17:58:31.937 & -29:12:17.86 & 20  & 5854$\pm$108 & 4.30$\pm$0.15 & 1.63$\pm$0.22 & -0.72$\pm$0.11 &  128.7	& 3  & 6.75$\pm$0.23  &  0.31$\pm$0.25  \\
2   & MOA-2009-BLG-133S  & 18:06:32.827 & -31:30:10.76 & 30  & 5529$\pm$73  & 4.30$\pm$0.14 & 1.17$\pm$0.18 & -0.69$\pm$0.07 &   91.6 	& 2  & 6.76$\pm$0.09  &  0.29$\pm$0.11 \\
3   & MOA-2009-BLG-475S  & 18:02:27.396 & -27:26:49.96 & 32  & 5836$\pm$189 & 4.40$\pm$0.27 & 1.35$\pm$0.37 & -0.52$\pm$0.20 &  137.8	& 4  & 6.86$\pm$0.21  &  0.22$\pm$0.29  \\
4   & MOA-2009-BLG-456S  & 17:48:56.377 & -34:13:32.34 & 81  & 5662$\pm$89  & 4.20$\pm$0.11 & 0.86$\pm$0.10 &  0.13$\pm$0.10 & -164.6 	& 5  & 7.46$\pm$0.16  &  0.17$\pm$0.19  \\
5   & MOA-2009-BLG-493S  & 17:55:46.018 & -28:48:25.81 & 61  & 5420$\pm$119 & 4.40$\pm$0.22 & 0.90$\pm$0.32 & -0.74$\pm$0.15 &  -14.5	& 3  & 6.68$\pm$0.12  &  0.26$\pm$0.19  \\
6   & MOA-2009-BLG-489S  & 17:57:46.563 & -28:38:57.80 & 108 & 5543$\pm$61  & 4.10$\pm$0.11 & 0.62$\pm$0.10 & -0.21$\pm$0.07 &   96.5	& 5  & 7.13$\pm$0.12  &  0.18$\pm$0.14  \\
7   & MOA-2010-BLG-049S  & 18:05:07.189 & -26:46:13.58 & 119 & 5694$\pm$61  & 4.10$\pm$0.12 & 1.02$\pm$0.12 & -0.40$\pm$0.07 & -116.7 	& 4  & 6.85$\pm$0.14  &  0.09$\pm$0.16  \\
8   & MOA-2010-BLG-037S  & 18:05:17.894 & -27:56:14.24 & 36  & 5732$\pm$109 & 3.90$\pm$0.20 & 1.52$\pm$0.19 &  0.55$\pm$0.20 &   -8.4 	& 4  & 7.58$\pm$0.06  & -0.13$\pm$0.21  \\
9   & MOA-2010-BLG-078S  & 17:52:01.723 & -30:24:25.49 & 85  & 5231$\pm$135 & 3.60$\pm$0.31 & 1.30$\pm$0.29 & -0.99$\pm$0.15 &   52.3 	& 3  & 6.60$\pm$0.15  &  0.43$\pm$0.21  \\
10  & MOA-2010-BLG-285S  & 17:56:48.131 & -30:00:40.03 & 122 & 6064$\pm$129 & 4.20$\pm$0.23 & 1.85$\pm$0.38 & -1.23$\pm$0.09 &   46.0 	& 3  & 6.16$\pm$0.14  &  0.23$\pm$0.17  \\
11  & MOA-2010-BLG-311S  & 18:08:49.837 & -25:57:05.00 & 82  & 5442$\pm$86  & 3.80$\pm$0.15 & 1.16$\pm$0.11 &  0.51$\pm$0.19 &   44.4 	& 4  & 7.62$\pm$0.11  & -0.05$\pm$0.22  \\
12  & MOA-2010-BLG-167S  & 18:11:27.495 & -29:41:02.04 & 65  & 5406$\pm$49  & 3.90$\pm$0.09 & 1.15$\pm$0.09 & -0.60$\pm$0.05 &  -79.4	& 5  & 6.86$\pm$0.11  &  0.30$\pm$0.12  \\
13  & MOA-2010-BLG-446S  & 18:07:04.236 & -28:03:56.45 & 82  & 6308$\pm$111 & 4.50$\pm$0.14 & 1.71$\pm$0.18 & -0.40$\pm$0.08 &   56.5 	& 4  & 6.94$\pm$0.05  &  0.18$\pm$0.09  \\
14  & MOA-2010-BLG-523S  & 17:57:08.919 & -29:44:58.96 & 171 & 5122$\pm$79  & 3.60$\pm$0.15 & 1.68$\pm$0.20 &  0.06$\pm$0.14 &   97.3 	& 3  & 7.33$\pm$0.12  &  0.11$\pm$0.18  \\
15  & MOA-2011-BLG-034S  & 18:10:12.088 & -27:32:02.18 & 65  & 5440$\pm$91  & 4.10$\pm$0.12 & 0.88$\pm$0.11 &  0.11$\pm$0.15 &  127.0	& 4  & 7.46$\pm$0.04  &  0.19$\pm$0.16  \\
16  & MOA-2011-BLG-058S  & 18:09:00.387 & -29:14:13.34 & 43  & 5256$\pm$100 & 4.00$\pm$0.15 & 0.71$\pm$0.12 &  0.37$\pm$0.25 & -139.7	& 4  & 7.57$\pm$0.09  &  0.04$\pm$0.27  \\
17  & MOA-2011-BLG-104   & 17:54:22.487 & -29:50:01.90 & 42  & 5900$\pm$173 & 4.10$\pm$0.28 & 1.30$\pm$0.59 & -0.85$\pm$0.15 &  197.3	& 3  & 6.72$\pm$0.17  &  0.41$\pm$0.23  \\
18  & MOA-2011-BLG-090   & 18:10:29.976 & -26:38:43.73 & 65  & 5367$\pm$49  & 4.10$\pm$0.09 & 0.87$\pm$0.09 & -0.26$\pm$0.05 &   48.2 	& 5  & 7.06$\pm$0.08  &  0.16$\pm$0.09  \\
19  & MOA-2011-BLG-174   & 17:57:20.571 & -30:22:47.50 & 71  & 6172$\pm$111 & 4.40$\pm$0.16 & 1.16$\pm$0.14 & -0.18$\pm$0.09 &  -24.0	& 5  & 7.16$\pm$0.06  &  0.18$\pm$0.11  \\
20  & MOA-2011-BLG-191   & 17:51:40.049 & -29:53:25.30 & 94  & 5382$\pm$92  & 3.80$\pm$0.13 & 0.57$\pm$0.11 &  0.26$\pm$0.14 &  134.0	& 4  & 7.51$\pm$0.08  &  0.09$\pm$0.16  \\
21  & OGLE-2011-BLG-1072 & 17:56:53.960 & -28:50:54.28 & 103 & 5515$\pm$89  & 3.90$\pm$0.13 & 1.11$\pm$0.10 &  0.36$\pm$0.12 &  -62.2	& 4  & 7.53$\pm$0.09  &  0.01$\pm$0.15  \\
22  & OGLE-2011-BLG-0950 & 17:57:16.478 & -32:39:57.64 & 29  & 6130$\pm$121 & 4.20$\pm$0.15 & 1.23$\pm$0.15 &  0.33$\pm$0.10 &   91.5 	& 3  & 7.48$\pm$0.02  &  -0.01$\pm$0.10  \\
23  & OGLE-2011-BLG-0969 & 18:09:41.300 & -31:11:03.73 & 33  & 6150$\pm$200 & 4.10$\pm$0.30 & 1.50$\pm$0.30 & -1.57$\pm$0.25 &  -53.0	& 3  & 5.80$\pm$0.07  &  0.21$\pm$0.26  \\
24  & OGLE-2011-BLG-1410 & 17:32:49.666 & -29:23:09.64 & 43  & 4831$\pm$108 & 3.30$\pm$0.24 & 0.62$\pm$0.15 &  0.22$\pm$0.37 &  -75.2	& 4  & 7.41$\pm$0.17  &  0.03$\pm$0.41  \\ 
25  & MOA-2011-BLG-455   & 18:04:45.608 & -28:35:43.33 & 50  & 4870$\pm$102 & 3.50$\pm$0.19 & 0.61$\pm$0.12 &  0.26$\pm$0.31 &   72.6 	& 3  & 7.51$\pm$0.14  &  0.09$\pm$0.34  \\
26  & OGLE-2012-BLG-0026 & 17:34:18.696 & -27:08:33.18 & 66  & 4815$\pm$145 & 3.40$\pm$0.28 & 0.62$\pm$0.14 &  0.50$\pm$0.44 &  132.2  & 3  & 7.64$\pm$0.08  &  -0.02$\pm$0.45  \\ 
27  & OGLE-2012-BLG-0211 & 18:10:10.945 & -25:01:40.19 & 49  & 5573$\pm$75  & 4.00$\pm$0.12 & 0.67$\pm$0.13 & -0.03$\pm$0.08 &  -17.7	& 4  & 7.26$\pm$0.12  &  0.13$\pm$0.14  \\
28  & OGLE-2012-BLG-0270 & 17:14:42.458 & -29:35:50.46 & 19  & 5914$\pm$145 & 4.30$\pm$0.22 & 1.33$\pm$0.22 & -0.84$\pm$0.13 & -123.7	& 3  & 6.69$\pm$0.18  &  0.37$\pm$0.22  \\
29  & MOA-2012-BLG-187   & 18:08:02.250 & -29:28:11.35 & 163 & 5892$\pm$94  & 4.20$\pm$0.14 & 1.60$\pm$0.18 & -0.74$\pm$0.08 &  -40.3	& 4  & 6.90$\pm$0.20  &  0.48$\pm$0.22  \\
30  & MOA-2012-BLG-202   & 18:12:34.693 & -25:03:01.33 & 90  & 4862$\pm$93  & 3.90$\pm$0.18 & 0.90$\pm$0.17 & -0.15$\pm$0.13 &   41.2 	& 3  & 7.27$\pm$0.14  &  0.26$\pm$0.19  \\
31  & MOA-2012-BLG-022   & 17:57:40.980 & -27:29:56.40 & 77  & 5827$\pm$111 & 4.30$\pm$0.12 & 0.89$\pm$0.11 &  0.42$\pm$0.09 &  -81.3	& 4  & 7.49$\pm$0.12  &  -0.09$\pm$0.15  \\ 
32  & OGLE-2012-BLG-0521 & 18:05:36.612 & -25:45:46.22 & 81  & 5013$\pm$84  & 3.70$\pm$0.14 & 1.02$\pm$0.11 &  0.09$\pm$0.15 &  -68.8	& 2  & 7.34$\pm$0.11  &  0.09$\pm$0.19  \\
33  & OGLE-2012-BLG-0563 & 18:05:57.630 & -27:42:44.57 & 56  & 5907$\pm$89  & 4.40$\pm$0.10 & 1.27$\pm$0.14 & -0.66$\pm$0.07 &  -66.2	& 3  & 6.83$\pm$0.10  &  0.33$\pm$0.12  \\
34  & OGLE-2012-BLG-0617 & 17:54:53.776 & -31:08:20.40 & 108 & 4924$\pm$71  & 3.70$\pm$0.16 & 1.06$\pm$0.11 & -0.14$\pm$0.09 &  -68.0	& 4  & 7.20$\pm$0.12  &  0.18$\pm$0.15  \\
35  & MOA-2012-BLG-291   & 18:02:43.062 & -28:23:03.16 & 38  & 5156$\pm$107 & 4.10$\pm$0.15 & 0.79$\pm$0.13 &  0.16$\pm$0.24 &   60.3 	& 4  & 7.49$\pm$0.14  &  0.17$\pm$0.28  \\
36  & MOA-2012-BLG-391   & 17:58:56.698 & -31:26:32.42 & 43  & 5505$\pm$76  & 3.90$\pm$0.14 & 1.12$\pm$0.16 & -0.24$\pm$0.07 &  -65.0	& 5  & 7.21$\pm$0.16  &  0.29$\pm$0.18  \\
37  & MOA-2012-BLG-410   & 18:10:22.548 & -25:10:17.80 & 30  & 5509$\pm$80  & 3.90$\pm$0.13 & 1.09$\pm$0.14 & -0.14$\pm$0.07 &   22.3 	& 4  & 7.18$\pm$0.19  &  0.16$\pm$0.20  \\
38  & OGLE-2012-BLG-1156 & 18:00:35.975 & -28:11:16.69 & 42  & 6200$\pm$200 & 4.20$\pm$0.30 & 1.50$\pm$0.30 & -1.89$\pm$0.25 &   72.6 	& 3  & 5.44$\pm$0.19  &  0.17$\pm$0.31  \\
39  & OGLE-2012-BLG-1217 & 18:10:16.808 & -27:38:01.32 & 57  & 5795$\pm$78  & 4.30$\pm$0.13 & 1.28$\pm$0.15 & -0.41$\pm$0.07 &  124.4	& 4  & 6.96$\pm$0.10  &  0.21$\pm$0.12  \\
40  & MOA-2012-BLG-532   & 17:58:41.253 & -30:02:13.24 & 27  & 5626$\pm$207 & 3.90$\pm$0.37 & 0.88$\pm$0.25 & -0.55$\pm$0.21 &   27.9 	& 3  & 6.84$\pm$0.14  &  0.23$\pm$0.25  \\
41  & OGLE-2012-BLG-1274 & 17:45:00.453 & -34:32:49.52 & 90  & 5733$\pm$51  & 4.10$\pm$0.07 & 1.23$\pm$0.07 &  0.07$\pm$0.04 &  -25.0	& 5  & 7.30$\pm$0.10  &  0.07$\pm$0.11  \\
42  & OGLE-2012-BLG-1534 & 18:00:46.444 & -28:01:02.39 & 92  & 5920$\pm$52  & 4.00$\pm$0.08 & 1.37$\pm$0.10 & -0.15$\pm$0.04 &  206.4	& 4  & 7.15$\pm$0.10  &  0.14$\pm$0.11  \\
43  & OGLE-2012-BLG-1526 & 18:09:43.092 & -28:48:47.52 & 30  & 5200$\pm$61  & 3.80$\pm$0.12 & 0.94$\pm$0.12 & -0.24$\pm$0.06 &  -87.1	& 4  & 7.16$\pm$0.11  &  0.24$\pm$0.13  \\
44  & MOA-2013-BLG-063S  & 17:45:13.377 & -33:29:50.86 & 22  & 6111$\pm$170 & 4.30$\pm$0.18 & 0.76$\pm$0.13 &  0.47$\pm$0.14 &   40.5 	& 3  & 7.54$\pm$0.12  &  -0.09$\pm$0.18  \\
45  & MOA-2013-BLG-068S  & 17:54:21.808 & -31:11:41.03 & 56  & 5312$\pm$48  & 3.70$\pm$0.10 & 1.05$\pm$0.09 & -0.27$\pm$0.05 &   68.7 	& 2  & 6.87$\pm$0.01  &  -0.02$\pm$0.05  \\ 
46  & OGLE-2013-BLG-0692 & 18:16:06.677 & -27:11:04.70 & 36  & 5015$\pm$88  & 3.60$\pm$0.13 & 0.79$\pm$0.10 &  0.15$\pm$0.19 &   74.2 	& 4  & 7.46$\pm$0.08  &  0.15$\pm$0.21  \\
47  & OGLE-2013-BLG-0446 & 18:06:56.268 & -31:39:27.83 & 183 & 5650$\pm$63  & 4.10$\pm$0.09 & 1.07$\pm$0.07 &  0.40$\pm$0.08 &   10.2 	& 4  & 7.53$\pm$0.03  &  -0.03$\pm$0.09  \\ 
48  & OGLE-2013-BLG-0835 & 17:52:59.295 & -29:05:59.46 & 149 & 4806$\pm$200 & 3.30$\pm$0.30 & 0.38$\pm$0.30 &  0.53$\pm$0.25 & -284.2	& 3  & 7.64$\pm$0.03  &  -0.05$\pm$0.25  \\ 
49  & MOA-2013-BLG-402S  & 18:03:00.110 & -29:54:25.13 & 36  & 4957$\pm$99  & 3.60$\pm$0.17 & 0.88$\pm$0.13 &  0.20$\pm$0.25 &  -44.2	& 3  & 7.13$\pm$0.21  &  -0.23$\pm$0.33  \\
50  & OGLE-2013-BLG-1114 & 17:54:24.460 & -28:56:31.63 & 50  & 6410$\pm$259 & 4.40$\pm$0.27 & 1.34$\pm$0.20 &  0.42$\pm$0.19 &   30.9 	& 4  & 7.50$\pm$0.21  &  -0.08$\pm$0.28  \\
51  & MOA-2013-BLG-524S  & 18:02:29.578 & -33:06:34.13 & 27  & 5932$\pm$289 & 4.40$\pm$0.37 & 1.95$\pm$0.68 & -1.01$\pm$0.26 &  -65.9	& 3  & 6.43$\pm$0.12  &  0.28$\pm$0.29  \\
52  & OGLE-2013-BLG-1147 & 18:08:39.105 & -26:40:43.32 & 46  & 5725$\pm$96  & 3.80$\pm$0.15 & 1.20$\pm$0.13 &  0.29$\pm$0.09 &   82.7 	& 4  & 7.49$\pm$0.04  &  0.04$\pm$0.09  \\
53  & MOA-2013-BLG-517S  & 18:13:36.248 & -27:43:22.76 & 42  & 6050$\pm$198 & 4.30$\pm$0.28 & 1.06$\pm$0.37 & -1.08$\pm$0.15 &  -35.9	& 3  & 6.30$\pm$0.12  &  0.22$\pm$0.19  \\
54  & OGLE-2013-BLG-1259 & 18:10:23.254 & -27:58:47.78 & 112 & 5299$\pm$64  & 3.90$\pm$0.12 & 1.06$\pm$0.08 & -0.21$\pm$0.06 &  -53.8	& 4  & 7.01$\pm$0.08  &  0.06$\pm$0.10  \\
55  & OGLE-2013-BLG-1015 & 17:52:48.240 & -35:00:53.14 & 91  & 5571$\pm$60  & 4.00$\pm$0.09 & 1.09$\pm$0.06 &  0.36$\pm$0.07 &   14.4 	& 5  & 7.48$\pm$0.05  &  -0.04$\pm$0.09  \\
56  & OGLE-2013-BLG-0911 & 17:55:31.855 & -29:15:15.66 & 86  & 5785$\pm$77  & 4.10$\pm$0.11 & 1.16$\pm$0.07 &  0.47$\pm$0.09 &  -46.8	& 4  & 7.61$\pm$0.03  &  -0.02$\pm$0.09  \\
57  & OGLE-2013-BLG-1793 & 17:54:04.707 & -29:38:05.75 & 93  & 5503$\pm$66  & 3.90$\pm$0.09 & 1.00$\pm$0.08 &  0.32$\pm$0.07 &  -10.8	& 4  & 7.50$\pm$0.13  &  0.02$\pm$0.15  \\
58  & OGLE-2013-BLG-1768 & 17:52:26.533 & -31:36:43.99 & 48  & 4842$\pm$200 & 3.50$\pm$0.30 & 1.36$\pm$0.30 &  0.11$\pm$0.25 &  180.9	& 3  & 7.38$\pm$0.15  &  0.11$\pm$0.29  \\
59  & OGLE-2013-BLG-1125 & 17:53:27.066 & -29:47:34.26 & 57  & 4739$\pm$86  & 3.00$\pm$0.18 & 1.17$\pm$0.10 & -0.12$\pm$0.07 & -232.7	& 3  & 7.12$\pm$0.26  &  0.08$\pm$0.27  \\
60  & MOA-2013-BLG-605S  & 17:58:42.828 & -29:23:55.21 & 52  & 4854$\pm$66  & 3.30$\pm$0.14 & 0.78$\pm$0.09 & -0.17$\pm$0.09 &   71.5 	& 4  & 7.07$\pm$0.10  &  0.08$\pm$0.14  \\
61  & OGLE-2013-BLG-1868 & 18:05:35.570 & -30:53:05.46 & 47  & 5732$\pm$78  & 4.10$\pm$0.11 & 1.30$\pm$0.10 &  0.38$\pm$0.07 &   98.2 	& 4  & 7.50$\pm$0.03  &  -0.04$\pm$0.08  \\
62  & OGLE-2013-BLG-1938 & 17:46:01.932 & -34:12:32.80 & 64  & 4921$\pm$60  & 3.30$\pm$0.13 & 1.00$\pm$0.11 & -0.41$\pm$0.09 &   17.4 	& 2  & 7.02$\pm$0.17  &  0.27$\pm$0.19  \\
63  & OGLE-2014-BLG-0157 & 17:58:16.726 & -33:46:19.96 & 44  & 4951$\pm$71  & 3.40$\pm$0.14 & 1.10$\pm$0.13 & -0.52$\pm$0.10 &    5.6 	& 4  & 6.97$\pm$0.09  &  0.33$\pm$0.14  \\
64  & MOA-2014-BLG-131S  & 17:59:02.586 & -31:01:54.16 & 51  & 5156$\pm$128 & 4.40$\pm$0.20 & 0.72$\pm$0.20 &  0.29$\pm$0.34 &   96.5 	& 3  & 7.48$\pm$0.09  &  0.03$\pm$0.35  \\
65  & OGLE-2014-BLG-0801 & 17:54:02.463 & -32:35:33.68 & 95  & 4720$\pm$150 & 2.90$\pm$0.28 & 0.71$\pm$0.11 &  0.55$\pm$0.34 &   31.4 	& 4  & 7.53$\pm$0.13  &  -0.18$\pm$0.36  \\
66  & OGLE-2014-BLG-0953 & 18:08:52.933 & -27:16:06.53 & 31  & 5447$\pm$109 & 3.90$\pm$0.20 & 1.12$\pm$0.19 & -0.13$\pm$0.12 &  -88.4	& 2  & 7.23$\pm$0.07  &  0.20$\pm$0.14  \\
67  & OGLE-2014-BLG-0987 & 18:16:53.621 & -25:31:31.84 & 49  & 5324$\pm$126 & 4.00$\pm$0.15 & 0.81$\pm$0.12 &  0.23$\pm$0.16 &   25.3 	& 3  & 7.57$\pm$0.17  &  0.18$\pm$0.23  \\
68  & OGLE-2014-BLG-1122 & 18:05:55.557 & -29:40:57.58 & 46  & 5552$\pm$63  & 4.20$\pm$0.12 & 1.07$\pm$0.13 & -0.32$\pm$0.06 &   -3.5 	& 4  & 7.01$\pm$0.18  &  0.17$\pm$0.19  \\
69  & OGLE-2014-BLG-1469 & 17:56:07.621 & -30:57:14.54 & 34  & 5517$\pm$80  & 3.80$\pm$0.14 & 1.04$\pm$0.13 & -0.70$\pm$0.08 &  288.3	& 4  & 6.77$\pm$0.07  &  0.31$\pm$0.11  \\
70  & OGLE-2014-BLG-1370 & 17:55:42.228 & -31:53:34.40 & 127 & 5272$\pm$111 & 4.10$\pm$0.16 & 0.69$\pm$0.13 &  0.43$\pm$0.29 & -158.1	& 4  & 7.59$\pm$0.10  &  0.00$\pm$0.31  \\ 
71  & OGLE-2014-BLG-1418 & 18:14:47.633 & -26:24:41.29 & 80  & 5481$\pm$85  & 4.10$\pm$0.12 & 0.98$\pm$0.09 &  0.44$\pm$0.14 & -128.0	& 4  & 7.57$\pm$0.15  &  -0.03$\pm$0.21  \\
72  & OGLE-2014-BLG-2040 & 17:55:59.488 & -31:16:35.00 & 21  & 5731$\pm$151 & 4.10$\pm$0.24 & 1.45$\pm$0.25 &  0.19$\pm$0.17 & -107.5	& 3  & 7.40$\pm$0.18  &  0.05$\pm$0.25  \\ 
73  & OGLE-2015-BLG-0078 & 17:55:30.540 & -27:57:38.56 & 62  & 5068$\pm$106 & 3.10$\pm$0.23 & 1.04$\pm$0.20 & -0.55$\pm$0.15 & -110.8	& 3  & 6.68$\pm$0.13  &  0.07$\pm$0.19  \\
74  & OGLE-2015-BLG-0159 & 17:43:38.626 & -35:05:13.92 & 27  & 5479$\pm$83  & 3.80$\pm$0.14 & 0.87$\pm$0.11 &  0.36$\pm$0.13 &  -74.0	& 4  & 7.52$\pm$0.12  &  0.00$\pm$0.18  \\

\end{longtable}
\end{landscape}
}

\longtab{
\begin{landscape}
\begin{longtable}{p{0.3cm}p{1.8cm}p{2cm}p{2cm}p{1.8cm}p{1.8cm}p{1.8cm}p{1.8cm}p{0.3cm}p{2cm}p{2cm}}
\caption{\label{thick} Summary of thick disk stars data. The target names and their coordinates are listed in columns 2-4. The atmospheric parameters estimated by \cite{2014A&A...562A..71B} follow in columns 5-8. The number of Sulfur line (N) used to estimate the NLTE Sulfur abundances obtained in this work are reported in the last two columns. The uncertainties on A(S) are the standard deviation.}\\
	\hline\hline
	&Object&RA&DEC&T$_{eff}$&log(g)&$\xi$&[Fe/H]&N&A(S)$_{\rm NLTE}$&[S/Fe]\\
	&&[deg]&[deg]&[K]&[cgs]&[kms$^{-1}$]&&&&\\
	\hline
	\endfirsthead
	
	\bfseries \tablename \thetable{ -- continued from previous page}\\
	\endhead

	\endfoot
	\hline\hline
	\endlastfoot
   1 & HIP13366 & 2.99319  & 11.36959   & 5856$\pm$50 & 4.26$\pm$0.08 & 1.04$\pm$0.07 & -0.69$\pm$0.04 & 2 & 6.65$\pm$0.02 &  0.17$\pm$0.04 \\  
   2 & HIP27128 & 6.28886  & -26.99131  & 5892$\pm$54 & 4.02$\pm$0.10 & 1.14$\pm$0.09 & -0.88$\pm$0.04 & 4 & 6.42$\pm$0.04 &  0.13$\pm$0.06 \\ 
   3 & HIP37853 & 16.39615 & -34.17491  & 5852$\pm$50 & 4.24$\pm$0.08 & 1.07$\pm$0.09 & -0.81$\pm$0.04 & 4 & 6.45$\pm$0.10 &  0.09$\pm$0.11 \\ 
   4 & HIP44075 & 34.68187 & -16.13187  & 5937$\pm$80 & 4.22$\pm$0.11 & 1.32$\pm$0.15 & -0.90$\pm$0.06 & 2 & 6.46$\pm$0.05 &  0.19$\pm$0.08 \\ 
   5 & HIP44896 & 37.20728 & -10.75756  & 5808$\pm$35 & 4.09$\pm$0.06 & 1.07$\pm$0.05 & -0.23$\pm$0.03 & 3 & 6.91$\pm$0.26 & -0.03$\pm$0.26 \\ 
   6 & HIP62607 & 92.43544 & 1.18602    & 5364$\pm$53 & 4.46$\pm$0.09 & 0.58$\pm$0.11 & -0.62$\pm$0.05 & 4 & 6.75$\pm$0.11 &  0.20$\pm$0.12 \\ 
   7 & HIP64426 & 98.08134 & 17.51720   & 5996$\pm$59 & 4.28$\pm$0.09 & 1.19$\pm$0.10 & -0.70$\pm$0.05 & 3 & 6.58$\pm$0.07 &  0.11$\pm$0.09 \\ 
   8 & HIP68464 & 10.25952 & -38.05084  & 6043$\pm$69 & 3.87$\pm$0.14 & 3.10$\pm$1.43 & -1.82$\pm$0.06 & 2 & 5.36$\pm$0.09 &  0.01$\pm$0.11 \\ 
   9 & HIP73650 & 25.82607 & -36.92599  & 5854$\pm$42 & 4.15$\pm$0.07 & 0.95$\pm$0.06 & -0.11$\pm$0.04 & 3 & 6.84$\pm$0.11 & -0.22$\pm$0.12 \\ 
  10 & HIP74033 & 26.94319 & 8.87892    & 5794$\pm$54 & 4.10$\pm$0.09 & 1.19$\pm$0.09 & -0.78$\pm$0.05 & 2 & 6.47$\pm$0.08 &  0.08$\pm$0.09 \\ 
  11 & HIP74067 & 27.05055 & -7.91344   & 5695$\pm$48 & 4.38$\pm$0.08 & 0.89$\pm$0.09 & -0.81$\pm$0.04 & 3 & 6.50$\pm$0.09 &  0.14$\pm$0.10 \\ 
  12 & HIP75181 & 30.44713 & -48.31851  & 5698$\pm$65 & 4.46$\pm$0.12 & 0.92$\pm$0.09 & -0.32$\pm$0.07 & 4 & 6.86$\pm$0.05 &  0.01$\pm$0.09 \\ 
  13 & HIP86013 & 63.67941 & 6.01433    & 5760$\pm$50 & 4.30$\pm$0.08 & 0.95$\pm$0.10 & -0.81$\pm$0.05 & 2 & 6.61$\pm$0.06 &  0.25$\pm$0.08 \\ 
  14 & HIP87101 & 66.94153 & -9.60507   & 5709$\pm$55 & 3.67$\pm$0.11 & 1.54$\pm$0.20 & -1.57$\pm$0.04 & 2 & 5.75$\pm$0.13 &  0.15$\pm$0.14 \\ 
  15 & HIP89554 & 74.10699 & -59.40311  & 6229$\pm$55 & 4.18$\pm$0.12 & 3.70$\pm$1.64 & -1.57$\pm$0.05 & 2 & 5.68$\pm$0.02 &  0.08$\pm$0.05 \\ 
  16 & HIP92781 & 83.59692 &  -4.60523  & 5765$\pm$64 & 4.33$\pm$0.10 & 0.81$\pm$0.13 & -0.69$\pm$0.06 & 2 & 6.68$\pm$0.01 &  0.20$\pm$0.06 \\ 
  17 & HIP95262 & 90.72369 & -32.91871  & 5759$\pm$46 & 4.01$\pm$0.07 & 0.87$\pm$0.07 & -0.51$\pm$0.04 & 2 & 6.74$\pm$0.16 &  0.08$\pm$0.16 \\ 
  18 & HIP98020 & 98.79034 &  10.74094  & 5379$\pm$78 & 4.51$\pm$0.11 & 1.03$\pm$0.42 & -1.69$\pm$0.08 & 2 & 6.12$\pm$0.20 &  0.64$\pm$0.22 \\ 
  19 & HIP98767 & 00.90695 &  29.89594  & 5572$\pm$92 & 4.47$\pm$0.12 & 0.83$\pm$0.11 &  0.26$\pm$0.16 & 4 & 7.39$\pm$0.10 & -0.04$\pm$0.19 \\ 
  20 & HIP10204 & 310.2063 & -18.79295  & 6009$\pm$54 & 4.13$\pm$0.08 & 1.41$\pm$0.13 & -1.05$\pm$0.04 & 4 & 6.34$\pm$0.18 &  0.22$\pm$0.18 \\ 
  21 & HIP10815 & 328.6939 & -73.43473  & 5613$\pm$75 & 4.30$\pm$0.09 & 0.94$\pm$0.07 &  0.03$\pm$0.08 & 3 & 7.24$\pm$0.08 &  0.04$\pm$0.11 \\ 
\end{longtable}
\end{landscape}
}

\longtab{
\begin{landscape}
\begin{longtable}{p{0.3cm}p{1.8cm}p{2cm}p{2cm}p{1.8cm}p{1.8cm}p{1.8cm}p{1.8cm}p{0.3cm}p{2cm}p{2cm}}
\caption{\label{uvespop} Summary of UVES POP stars data. The target names and their coordinates are listed in columns 2-4. The atmospheric parameters estimated by \cite{2014A&A...562A..71B} follow in columns 5-8. The number of Sulfur line (N) used to estimate the NLTE Sulfur abundances obtained in this work are reported in the last two columns. The uncertainties on A(S) are the standard deviation.}\\
	\hline\hline
	&Object&RA&DEC&T$_{eff}$&log(g)&$\xi$&[Fe/H]&N&A(S)$_{\rm NLTE}$&[S/Fe]\\
	&&[deg]&[deg]&[K]&[cgs]&[kms$^{-1}$]&&&&\\
	\hline
	\endfirsthead
	
	\bfseries \tablename \thetable{ -- continued from previous page}\\
	\endhead

	\endfoot
	\hline\hline
	\endlastfoot
1   & HD739	  & 002.93342 &     -35.13312 &	  6547$\pm$76 & 4.14$\pm$0.10 & 1.38$\pm$0.06 & -0.02$\pm$0.05 & 3 & 7.09$\pm$0.03 &  -0.05$\pm$0.06  \\
2   & HD1581	  & 005.01775 &     -64.87479 &   5932$\pm$61 & 4.33$\pm$0.09 & 1.00$\pm$0.07 & -0.19$\pm$0.06 & 2 & 6.95$\pm$0.00 &  -0.02$\pm$0.06  \\	
3   & HD2151	  & 006.43779 &     -77.25425 &	  5852$\pm$43 & 3.91$\pm$0.07 & 1.27$\pm$0.07 & -0.07$\pm$0.05 & 3 & 7.04$\pm$0.02 &  -0.05$\pm$0.05  \\	
4   & HD3158	  & 008.61597 &     -52.37309 &	  6499$\pm$84 & 4.22$\pm$0.10 & 1.49$\pm$0.10 &  0.06$\pm$0.06 & 2 & 7.12$\pm$0.00 &  -0.10$\pm$0.06  \\	
5   & HD10647	  & 025.62215 &     -53.74083 &	  6219$\pm$60 & 4.41$\pm$0.06 & 1.07$\pm$0.06 &  0.05$\pm$0.05 & 4 & 7.15$\pm$0.07 &  -0.06$\pm$0.09  \\	
6   & HD14802	  & 035.63585 &     -23.81680 &	  5917$\pm$53 & 3.96$\pm$0.07 & 1.26$\pm$0.05 &  0.00$\pm$0.05 & 5 & 7.03$\pm$0.11 &  -0.13$\pm$0.12  \\	
7   & HD16673	  & 040.05176 &     -09.45287 &	  6383$\pm$59 & 4.41$\pm$0.07 & 1.24$\pm$0.06 &  0.09$\pm$0.05 & 4 & 7.17$\pm$0.09 &  -0.08$\pm$0.10  \\ 	
8   & HD20010	  & 048.01886 &     -28.98762 &	  6199$\pm$50 & 3.92$\pm$0.08 & 1.40$\pm$0.06 & -0.18$\pm$0.04 & 2 & 6.94$\pm$0.02 &  -0.04$\pm$0.05  \\ 	
9   & HD20807	  & 049.55341 &     -62.50637 &	  5826$\pm$47 & 4.41$\pm$0.08 & 0.93$\pm$0.07 & -0.24$\pm$0.05 & 5 & 7.09$\pm$0.15 &  0.17$\pm$0.16  \\	
10  & HD22484	  & 054.21727 &     +00.39959 &	  6036$\pm$47 & 4.08$\pm$0.07 & 1.30$\pm$0.05 & -0.03$\pm$0.04 & 3 & 7.12$\pm$0.05 &  -0.01$\pm$0.06 \\ 	
11  & HD30562	  & 072.15160 &     -05.67404 &	  5923$\pm$74 & 4.06$\pm$0.09 & 1.31$\pm$0.05 &  0.26$\pm$0.07 & 4 & 7.37$\pm$0.08 &  -0.05$\pm$0.11  \\ 	
12  & HD33256	  & 077.18208 &     -04.45621 &   6406$\pm$64 & 3.94$\pm$0.10 & 1.56$\pm$0.07 & -0.24$\pm$0.05 & 4 & 6.82$\pm$0.05 &  -0.1$\pm$0.07  \\	
13  & HD38393     & 086.11580 &     -22.44838 &   6323$\pm$77 & 4.16$\pm$0.10 & 1.39$\pm$0.08 & -0.02$\pm$0.06 & 4 & 7.06$\pm$0.02 &  -0.08$\pm$0.06  \\	
14  & HD43318     & 093.89277 &     -00.51219 &	  6337$\pm$51 & 3.89$\pm$0.08 & 1.47$\pm$0.06 & -0.08$\pm$0.04 & 4 & 7.04$\pm$0.06 &  -0.04$\pm$0.07  \\	
15  & HD45067     & 096.31894 &     -00.94588 &	  6042$\pm$41 & 3.89$\pm$0.06 & 1.34$\pm$0.05 & -0.03$\pm$0.03 & 4 & 7.05$\pm$0.01 &  -0.08$\pm$0.03  \\ 	
16  & HD59468     & 111.85610 &     -51.40260 &	  5564$\pm$53 & 4.38$\pm$0.07 & 0.88$\pm$0.07 &  0.03$\pm$0.07 & 5 & 7.20$\pm$0.05 &  0.01$\pm$0.09  \\	
17  & HD59967     & 112.67713 &     -37.33936 &	  5859$\pm$57 & 4.52$\pm$0.08 & 1.03$\pm$0.06 &  0.00$\pm$0.06 & 5 & 7.18$\pm$0.09 &  0.02$\pm$0.11  \\	
18  & HD91889     & 159.13492 &     -12.23012 &   6089$\pm$71 & 4.12$\pm$0.11 & 1.14$\pm$0.07 & -0.18$\pm$0.06 & 4 & 6.77$\pm$0.08 &  -0.21$\pm$0.10  \\
19  & HD100623    & 173.62286 &     -32.83134 &   5131$\pm$65 & 4.50$\pm$0.11 & 0.60$\pm$0.13 & -0.40$\pm$0.07 & 4 & 6.72$\pm$0.10 &  -0.04$\pm$0.12  \\	
20  & HD105113    & 181.52165 &     -32.96123 &   5909$\pm$50 & 3.83$\pm$0.07 & 1.30$\pm$0.05 & -0.10$\pm$0.04 & 4 & 7.05$\pm$0.16 &  -0.01$\pm$0.17  \\	
21  & HD115383    & 199.19382 &     +09.42416 &	  6185$\pm$60 & 4.28$\pm$0.07 & 1.41$\pm$0.08 &  0.25$\pm$0.06 & 5 & 7.26$\pm$0.13 &  -0.15$\pm$0.14  \\	
22  & HD115617    & 199.60131 &     -18.31120 &   5469$\pm$75 & 4.36$\pm$0.08 & 0.85$\pm$0.09 & -0.03$\pm$0.09 & 4 & 7.33$\pm$0.19 &  0.20$\pm$0.21  \\	
23  & HD128167    & 218.67007 &     +29.74513 &   6958$\pm$80 & 4.27$\pm$0.13 & 1.60$\pm$0.11 & -0.24$\pm$0.05 & 2 & 6.74$\pm$0.00 &  -0.18$\pm$0.05  \\	
24  & HD136351    & 230.53446 &     -47.92779 &	  6447$\pm$80 & 3.87$\pm$0.10 & 1.60$\pm$0.07 &  0.17$\pm$0.05 & 4 & 7.24$\pm$0.09 &  -0.09$\pm$0.10  \\
25  & HD140901    & 236.87125 &     -37.91631 &   5582$\pm$68 & 4.45$\pm$0.08 & 0.94$\pm$0.10 &  0.11$\pm$0.11 & 4 & 7.28$\pm$0.03 &  0.01$\pm$0.11  \\	
26  & HD149661    & 249.08937 &     -02.32459 &	  5216$\pm$88 & 4.59$\pm$0.12 & 0.98$\pm$0.12 & -0.01$\pm$0.14 & 4 & 7.24$\pm$0.08 &  0.09$\pm$0.16  \\ 	
27  & HD152311    & 253.35499 &     -20.41578 &	  5703$\pm$56 & 3.93$\pm$0.07 & 1.20$\pm$0.05 &  0.19$\pm$0.06 & 5 & 7.20$\pm$0.08 &  -0.15$\pm$0.10  \\ 	
28  & HD189340    & 299.94700 &     -09.95823 &   5867$\pm$59 & 4.32$\pm$0.09 & 1.10$\pm$0.08 & -0.08$\pm$0.06 & 4 & 7.12$\pm$0.09 &  0.04$\pm$0.11  \\	
29  & HD211415    & 334.56506 &     -53.62707 &   5825$\pm$47 & 4.38$\pm$0.07 & 0.99$\pm$0.06 & -0.23$\pm$0.05 & 5 & 6.94$\pm$0.06 &  0.01$\pm$0.08  \\ 	
30  & HD216435    & 343.40805 &     -48.59828 &   6024$\pm$57 & 4.10$\pm$0.07 & 1.39$\pm$0.06 &  0.26$\pm$0.05 & 3 & 7.36$\pm$0.12 &  -0.06$\pm$0.13  \\
\end{longtable}
\end{landscape}
}

\begin{table}
	\centering
	\caption{Atomic parameters of the sulfur lines.}
	\label{lines}
	\small
	\centering
	\begin{tabular}{ccccc}	
		\hline\hline
		Wavelength & Mult. & Transition & log $gf$ & $\chi_{\rm    low}$\\
		$[$nm$]$& & & & $[$eV$]$\\
		\hline
		675.7171 & 8 &$^5$P$_3-^5$D$_4^{\rm o}$ &-0.35&7.869\\		
		869.4626 & 6 &$^5$P$_3-^5$D$_4^{\rm o}$ &0.05&7.869\\
		921.2863 & 1 &$^5$S$_2^{\rm o}-^5$P$_3$ &0.40 &6.525\\
		922.8093 & 1 &$^5$S$_2^{\rm o}-^5$P$_2$ &0.25&6.525\\
		923.7538 & 1 &$^5$S$_2^{\rm o}-^5$P$_1$ &0.03&6.525\\
		\hline
		\hline
	\end{tabular}
\end{table}

\begin{acknowledgements}
The authors are grateful to the anonymus referee for providing helpful comments and suggestions, which have improved the content of this paper.

Support for the author FL is provided by CONICYT-PFCHA/Doctorado Nacional año 2020-folio 21200677.

EC and PB acknowledge support from the French
National Research Agency (ANR) funded project ``Pristine''
(ANR-18-CE31-0017).

\end{acknowledgements}

\newpage
\bibliographystyle{aa.bst}
\bibliography{biblio}
\end{document}